\documentclass[11pt]{article}

\usepackage[utf8]{inputenc}
\usepackage[english]{babel}

\usepackage{mathpazo}
\usepackage{amsmath}
\usepackage{amsthm}
\usepackage{amssymb,bm}
\usepackage{mathrsfs}
\usepackage{stmaryrd}
\usepackage{multicol}
\usepackage{wrapfig}

\usepackage[unicode=true, 
            colorlinks=true, 
            linkcolor=blue,
            citecolor=blue,
            urlcolor=blue]{hyperref}
\usepackage[authoryear]{natbib}

\usepackage{amssymb}
\usepackage{multirow}
\usepackage{amsmath}
\usepackage{enumitem}  
\usepackage{pifont}
\usepackage{footmisc}
\usepackage{graphicx,wrapfig,lipsum}

\usepackage[left=2.7cm,right=2.7cm,top=2.cm,bottom=2.cm]{geometry}

\usepackage{comment}
\newcommand{\ARR}{{\operatorname{ARR}}}

\newcommand{\cmark}{\color{black}\ding{51}}
\newcommand{\xmark}{\color{black}\ding{55}}
\theoremstyle{plain}

\theoremstyle{definition}
\usepackage[]{graphicx}
\chardef\bslash=`\\ % p. 424, TeXbook

\usepackage{cleveref}
\crefname{equation}{Eq.}{Eqs.}
\crefname{figure}{Fig.}{Figs.}
\crefname{table}{Table}{Tables}
\crefname{section}{Section}{Secs.}
\crefname{appendix}{Appendix}{Appendixes}

\usepackage{footmisc}
\usepackage{authblk}

\title{Subgroup analysis methods for time-to-event outcomes in heterogeneous randomized controlled trials}

\author{
  Valentine Perrin\footnote{Corresponding author: {\sf{e-mail: valentine.perrin@owkin.com}}, \newline ${\phantom{aal}}^{\dag}$ Equal contributions, \newline ${\phantom{aal}}^{\ddag}$ Owkin Inc, New York, USA.}\textsuperscript{  \hspace{0.05cm},}\textsuperscript{\dag, \ddag},
  Nathan Noiry\textsuperscript{\dag, \ddag},
  Nicolas Loiseau\textsuperscript{\ddag},
  Alex Nowak\textsuperscript{\ddag}
}

\begin{document}

\maketitle

\begin{abstract}
Non-significant randomized control trials can hide subgroups of good responders to experimental drugs, thus hindering subsequent development. Identifying such heterogeneous treatment effects is key for precision medicine and many post-hoc analysis methods have been developed for that purpose. While several benchmarks have been carried out to identify the strengths and weaknesses of these methods, notably for binary and continuous endpoints, similar systematic empirical evaluation of subgroup analysis for time-to-event endpoints are lacking. This work aims to fill this gap by evaluating several subgroup analysis algorithms in the context of time-to-event outcomes, by means of three different research questions:
Is there heterogeneity? What are the biomarkers responsible for such heterogeneity? Who are the good responders to treatment?
In this context, we propose a new synthetic and semi-synthetic data generation process that allows one to explore a wide range of heterogeneity scenarios with precise control on the level of heterogeneity. 
We provide the open source Python package \href{https://github.com/owkin/hte}{\texttt{hte}}, available on Github, containing our generation process and our comprehensive benchmark framework. We hope this package will be useful to the research community for future investigations of heterogeneity of treatment effects and subgroup analysis methods benchmarking.
\end{abstract}

\maketitle

\section{Introduction}

Randomized controlled trials (RCTs) are the gold standard for causal inference in drug clinical development. The goal of RCTs is to perform an averaged efficacy analysis of a drug over a predefined population of patients by treatment randomization. However, the averaged analysis may conceal high variability in treatment response in terms of patient profiles, in which case not all patients do benefit from the treatment. Dealing with this heterogeneity of treatment effect (HTE) is the very basis of precision medicine. 

Finding subgroups of the population with homogeneous treatment response within an RCT has many applications, such as better identification of patient populations with optimal benefit from a drug, identification of super-responders and safety profiling, amongst others -- see the surveys of \cite{grouin2005}, \cite{tanniou2016} or \cite{amatya2021subgroup} for more details. Approval of olaparib in combination with bevacizumab for advanced homologous recombination deficiency (HRD)-positive epithelial ovarian cancer is a notable example of approval backed up by subgroup analysis. Indeed, post-hoc analysis of PAOLA-1 trial \citep{ray2019olaparib} results showed that the positive significant treatment effect was solely driven by the HRD$+$ subgroup while the HRD$-$ subgroup even showed negative treatment effect for survival \citep{arora2021fda}. 

Subgroup analysis encompasses two distinct approaches. The first approach, known as confirmatory subgroup analysis, relies on pre-specified statistical analysis with a primary emphasis on controlling false discovery rates. In contrast, exploratory subgroup analyses are more flexible and mainly aim at generating new hypotheses by discovering previously unknown patterns within the data. Despite inflated false discovery rates \citep{brookes2004}, the European medicines agency (EMA) underlined this exploratory aspect and made the point that absence of pre-specification could not be taken as a direct argument for a lack of results credibility \citep{ema2019}. Indeed, although results of post-hoc investigations require further validation, they can nevertheless inform future strategic decisions within the drug discovery pipeline of pharmaceutical groups.

In this article, we focus on the rather common situation where the resulting effect of a trial is non-significant, but practitioners have good reasons to believe good responders exist in the population. In this context, we benchmark several subgroup analysis algorithms to gain insights into their usefulness for HTE investigations. We consider time-to-event endpoints which are some of the most common primary endpoints in RCTs. This choice is also motivated by the fact that several benchmarks of HTE methodologies have been conducted on binary and continuous endpoints, while our understanding of time-to-event endpoints in this context remains limited -- see \cref{sec:comparative}.

The question of subgroup identification is not new and many methods have been developed, as discussed later in \cref{sec:comparative}. Although different frameworks exist, they all consist of finding homogeneous regions among patients. We selected 9 methods to obtain a representative panel in our benchmark. Some of them have only been introduced theoretically in the case of time-to-event data, others are only available in R \citep{rsoftware}. In order to ease their adoption, we have implemented all of them in Python \citep{python}.

Comparing subgroup analysis algorithms implies {\it (i)} having access to a form of ground truth and {\it (ii)} relying on relevant evaluation metrics to quantify the distance to this ground truth. We elaborate on these two points in the next two paragraphs.

Retrospective data of RCTs do not contain ground truth subgroups responsible for heterogeneity. Indeed, even if a subgroup of good responders has been identified and validated in RCTs, the existence of this subgroup does not rule out the existence of another subgroup not yet identified. It is therefore useful to rely on synthetic datasets for benchmarking subgroup analysis methods: despite their inevitable simplification of biological complexity, synthetic datasets provide a notion of ground truth which is key for an objective evaluation of the subgroup analysis algorithms. We introduce a synthetic time-to-event data generation process that provides precise control on the level of treatment effect heterogeneity. The latter is a Cox model with an interaction term between the treatment and a functional of the covariates that determines whether a patient is a good or bad responder. In our synthetic experiments, we take Gaussian covariates, and also investigate more complex laws and covariance structure through semi-synthetic experiments, where the covariates follow the empirical distribution of a real world dataset. This is further discussed in \cref{sec:dgp}.

Following the approach of \cite{sun2022}, we define appropriate metrics based on research questions of interest that subgroup analysis methods aim at answering. Notably, we are interested in {\it (i)} the existence of treatment effect heterogeneity; {\it (ii)} the identification of variables responsible for the heterogeneity; {\it (iii)} the identification of good and bad responders. Each of these questions should be evaluated using relevant metrics, discussed in more detail in \cref{sec:metrics}. 

With the hope to support future HTE research avenues, we open source a Python package, \href{https://github.com/owkin/hte}{\texttt{hte}}, containing the data generation process, the implemented methods, as well as the evaluation framework (metrics and processing of results) introduced in this paper. We propose a modular code, with the possibility to generate synthetic and semi-synthetic data independently of the benchmark, or to use the benchmark framework on other subgroup analysis methods of the user's interest.   

\medskip

\noindent {\bf Organization of the paper.} In \cref{sec:comparative}, we review existing comparative synthetic data studies. Then, in \cref{sec:dgp}, we introduce the data generation process we developed to investigate a wide range of heterogeneity scenarios in a time-to-event context. In \cref{sec:metrics}, we present the research questions of interest and their corresponding evaluation metrics, before turning to the methods under investigation that we introduce in \cref{sec:methods}. Finally, we present the results of our simulation study in \cref{sec:results} and a final discussion in \cref{sec:discussion}. 

\section{Related work and contributions}
\label{sec:comparative}

Many methods have been developed to perform subgroup analysis, with broad applications that go beyond biomedical sciences such as in social and economic sciences \citep{imbens2015causal}. As previously stated, subgroup identification mostly consists of finding groups of homogeneous samples. Roughly speaking, we can distinguish two main communities. The biostatistics community, which is mainly interested in finding subgroups of patients in randomized control trials, and the machine learning community, with a broader area of interest that includes A/B testing experiments for industrial applications -- see the works of \cite{larsen2022statistical} or \cite{xie2018false}. Of course, these two approaches are not always clearly distinct, but we think this dichotomy helps classifying previous works.

\medskip

\noindent {\bf Biostatistics approaches.} Safety issues, and clinical success stakes of patients subtyping have prompted the development of methods that {\it (i)} control the false positive rates and {\it (ii)} are interpretable by design. These requirements can limit the predictive power of the methods which may not be able to fit data complexity. Apart from the traditional univariate analysis where each covariate is processed separately, methods from the biostatistics community tend to rely on recursive explorations of the data to find relevant subgroups of patients. We can distinguish tree-based and rule-based methods. Tree-based algorithms recursively partition samples by growing tree-like structures, where the subgroups of patients correspond to the tree leaves and are characterized by their path from the root node. We identified five existing tree-based methods: model-based partitioning \citep{seibold2016}, interaction tree \citep{su2008}, SIDES \citep{lipkovich2011}, GUIDE \citep{loh2002}, and STIMA \citep{dusseldorp2010}. Rule-based methods such as AIM \citep{tian2011}, PRIM \citep{chen2015}, SeqBT \citep{huang2017}, and ARDP \citep{leblanc2005}, partition samples by fine-tuning a subgroup rule through many iterations and using the final rule to define subgroups of different treatment effects. 

\medskip

\noindent {\bf Machine learning approaches.} On the other hand, machine learning algorithms are able to fit complex functions, at the price of a loss of interpretability. Indeed, machine learning models are algorithms with potentially many parameters; this can bring huge flexibility to estimate complex distributions, but the interpretation of each parameter of the model is highly challenging compared to a simple linear regression. The most representative methods for HTE analysis estimate the conditional averaged treatment effect (CATE) and leverages it for subgroup identification, see for instance \cite{chernozhukov2018, wager2018} or the FindIt procedure \citep{imai2013}. The recent work of \cite{xu2023treatment} provides a comparison of different CATE learning approaches in the context of time-to-event outcomes. Some other methods look for the treatment allocation rule that maximizes the expected reward, i.e. the difference of treatment effect: OWL \citep{zhao2012} and ROWSi \citep{xu2015}. Of note, let us emphasize that the interaction tree \citep{su2008} method could also be classified as a machine learning approach, due to its similarity with the famous CART algorithm \cite{loh2011classification} and its high number of hyperparameters.

\medskip

\noindent {\bf Benchmarking subgroup analysis methods.} The complexity of the subgroup identification question and the ever-increasing number of methods have subsequently led to the creation of numerous benchmarks in many different subgroup settings. In this paper, we do not aim to perform an extensive review of existing benchmarks, but rather discuss the different settings evaluated.  Most reviewed benchmark papers \citep{ternes2017, huang2017, alemayehu2018, sechidis2018, huber2019, loh2019} involve synthetic data, with the exception of \cite{sun2022} that focuses on semi-synthetic data. The existing work is mostly focused on studying heterogeneity in the context of binary endpoints \citep{sechidis2018, loh2019, sun2022} or of continuous endpoints \citep{huang2017, alemayehu2018, huber2019, sun2022}.
Apart from \cite{sechidis2018} and \cite{ternes2017} who studied dimensions up to 400 and 1000 respectively, most benchmarks are performed in small dimensional settings between 1 and 30 covariates \citep{huang2017, alemayehu2018, huber2019, loh2019, sun2022}. With the exception of \cite{ternes2017} we have not identified any benchmark paper of subgroup analysis methods in the context of time-to-event endpoints. In this paper, the authors focus on the identification of biomarkers that are responsible for heterogeneity in the sense of an interaction with the treatment. To evaluate several machine-learning based selection methods, they rely on a synthetic data generating process: a Cox model with interaction terms. They make the number of variables interacting with the treatment vary to explore different heterogeneity scenarios. The main difference with our data generation process is that we provide a way to select the coefficients in order to achieve pre-specified heterogeneity levels, in terms of Absolute Risk Reduction (this notion is introduced in Section \ref{sec:dgp}). In addition, our benchmark extends beyond variable selection to encompass the investigation of heterogeneity existence and patient selection: this makes us look at different subgroup analysis methods.

\medskip

\noindent {\bf Contributions.} Our benchmark focuses on the time-to-event setting, which is a highly prevalent outcome in clinical trials. We extend existing work by {\it (i)} focusing on the most common research questions tackled by subgroup analysis; {\it (ii)} adapting existing methods to answer all these questions; and {\it (iii)} providing a data generation process tailored at benchmarking in this time-to-event setting. More specifically, these are our main contributions:
\begin{enumerate}
    \item {\bf Data generation process.} We introduce a new time-to-event data generating procedure allowing to define complex subgroups of good/bad responders and to precisely control the treatment responses within the subgroups (\cref{sec:dgp}). This allows us to explore a continuum of heterogeneity scenarios. In addition, we explore small ($p=20$) and high ($p=1000$) dimensional settings, including synthetic and semi-synthetic generations.
    \item {\bf Research questions.} Heterogeneity of treatment effect is a broad subject that is not restricted to biomarkers identification. We benchmark several subgroup analysis methods through the lens of three research questions and their associated metrics: existence of heterogeneity, biomarkers identification and good responders identification -- see \cref{sec:metrics}.
    \item {\bf Subgroup analysis methods.} We selected 9 methods in our benchmark to cover various approaches that can answer the above questions. Some of these methods were only available in R, others not yet implemented. We implemented all of them in Python. See \cref{sec:methods} and Appendix \ref{appendix:methods} for details.
    \item {\bf Python package.} We open source a \href{https://github.com/owkin/hte}{Python package} containing the benchmark framework we developed. This package contains two distinct modules that can be used independently: {\it (i)} a module containing the data generating process we introduce in this paper, allowing for the generation of synthetic and semi-synthetic data with precise HTE control and modular subgroups; {\it (ii)} the benchmark module with all 9 subgroup analysis methods implemented, associated evaluation metrics, results processing, and the possibility to include any method suiting the framework. 
\end{enumerate}

\section{Data generation process}
\label{sec:dgp}

In this section we introduce the data generation process (DGP) we developed for the time-to-event subgroup analysis methods benchmark. This will in particular allow us to clarify the null and alternative hypotheses we are working with. We first remind the necessary basics of survival analysis and the statistical framework for treatment effect estimation. 

\medskip

\noindent {\bf Notations.} In an RCT with a time-to-event primary endpoint, one is interested in the probability of occurrence of an event of interest (e.g. death) after treatment induction. We denote by $W$ the treatment variable, which is equal to $0$ if the patient belongs to the control arm and to $1$ if they are treated. Since we are in an RCT, $W$ follows a Bernoulli law with parameter $1/2$.

\medskip

\noindent {\bf Time-to-event framework.} A patient is described by a series of variables $\bf X$ measured at baseline (e.g. the age, the expression of a gene, etc). The random variable $\bf X$ takes values in $\mathbb{R}^p$, where the dimension $p \geq 1$ is the number of variables describing a patient. Denoting by $T$ the (random) time at which the event occurs, one is interested in the survival function of a patient $S(\cdot \, | \, \mathbf{X})$. Computed at a specific time $t \geq 0$, the survival function is the probability that $T$ exceeds $t$, given the knowledge of the patient description at baseline: $S(t \, | \, \mathbf{X}) = \mathbb{P}(T \geq t \, | \, \mathbf{X})$. 

\medskip

\begin{wrapfigure}{r}{5.1cm}
\includegraphics[width=5.3cm]{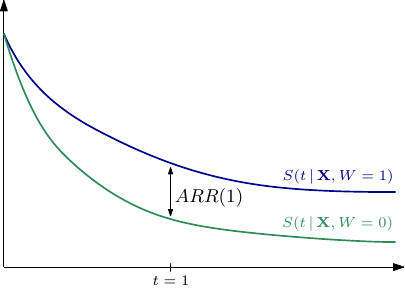}
\caption{Illustration of the Absolute Risk Reduction.}\label{fig:def-arr}
\end{wrapfigure}

\noindent {\bf Measuring the efficacy of a treatment.} One of the measures used to assess the efficacy of a treatment in an RCT context is the absolute risk reduction (ARR) at timepoint $t$:
\[ \ARR(t) := S(t \, | \, \mathbf{X}, W=1) - S(t \, | \, \mathbf{X}, W=0).  \]
This measure accounts for the benefit, after a period of time $t$, of having received the treatment. Visually, it is the vertical gap between the two survival curves at time $t$ -- see \cref{fig:def-arr} for an illustration. 
\noindent The $\ARR$ is collapsible, meaning that it behaves well under the average operation. Let us give an example for clarity. Say we are interested in two subgroups of the population, depending on the sex of the patients $\xi = 0$ for male and $\xi = 1$ for female. Denoting $\ARR_0$ (resp. $\ARR_1$) the $\ARR$ of male patients (resp. female patients), the law of total probability leads to $\ARR = \mathbb{P}(\xi=0)\ARR_0(t) + \mathbb{P}(\xi=1)\ARR_1(t)$. In short, this is saying that $\ARR(t) = \mathbb{E}_\xi[\ARR(t \, | \, \xi)]$ and this holds for general random variables $\xi$. As such the $\ARR$ is amenable to subgroup analysis, which is not the case of the canonical hazard ratio that suffers from a lack of causal interpretation, see for instance \cite{hernan2010hazards} or \cite{aalen2015does}. We also refer to \cite{daniel2021making} for general considerations about the pitfalls of non-collapsibility in the context of heterogeneity investigations. Note that another collapsible measure is the restricted mean survival time (RMST), which is the expected value of time to event before a predefined time point, but we only focus on the ARR in this paper. For our computations, we set $t=1$.

\medskip

\noindent {\bf Our synthetic model.} Our synthetic data generating process consists of using a Cox parametric model for the survival function $S$, with a treatment coefficient that varies depending on the nature of the covariates $\bf X$. We denote by $h$ the hazard function defined by $h(t) = \lim_{h \rightarrow 0 } [S(t+h) - S(t)]/h$. The hazard is the probability that an individual who is under observation at a time t has an event at that time. For our data generating process, we take:
\begin{equation}\label{eq:our_synthetic_model}
h(t \, | \, \mathbf{X}, W) = h_0(t) \exp\left( \beta_0 W + (\beta_1 - \beta_0) G(\mathbf{X}) W + \gamma^T \mathbf{X} \right),
\end{equation}
where $h_0(t)$ is the baseline hazard and $\gamma \in \mathbb{R}^p$ a prognostic vector. Moreover, $G: \mathbb{R}^p \rightarrow \{0, 1\}$ is the subgroup function which allocates a patient to the subgroup of bad ($G=0$) or good ($G=1$) responders, which have respective log-hazard ratio given by $\beta_0$ and $\beta_1$. As a side node, a biomarker is {\it prognostic} when it affects the outcome irrespective of the treatment, while it is said to be {\it predictive} when it modifies the treatment effect. In our case, the variable $G(\mathbf{X})$ is predictive. The choice of the function $G$ allows us to control the {\it nature} of the heterogeneity, while the coefficients $\beta_0$ and $\beta_1$ allow us to control the {\it level} of heterogeneity. Note that the law of $\bf X$ is also an implicit parameter of our generation process. Note also that our DGP makes a Proportional Hazard assumption between good and bad responders groups. 

\medskip 

\begin{figure}[htb]
\begin{center}
\includegraphics[scale=0.8]{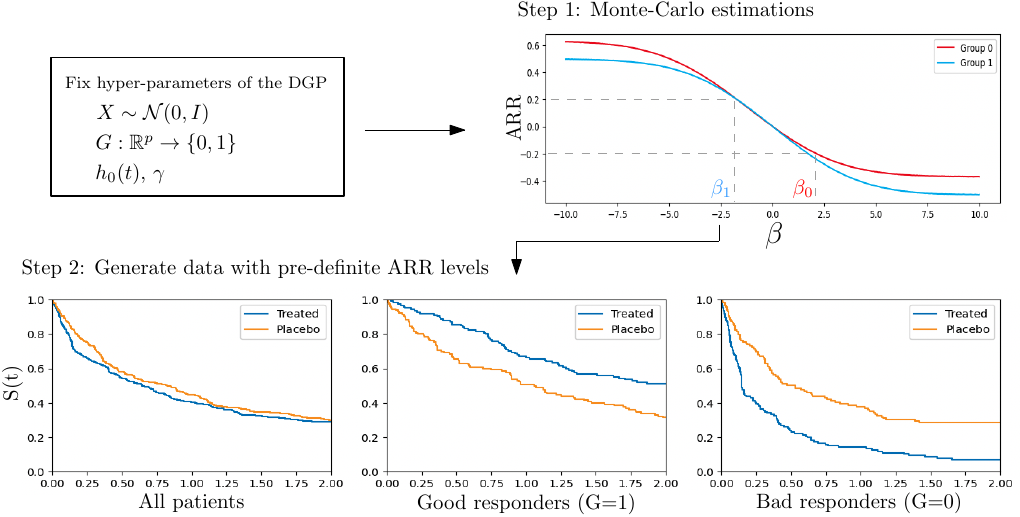}
\caption{For given values of the hyperparameters of the data generating process, step 1 consists of estimating the $\ARR$ values in each subgroup (Group $G=0$ and Group $G=1$) in function of $\beta$. Then for previously defined $\ARR$ levels in each subgroup (here $-0.2$ and $0.2$), we are able to select the corresponding parameters $\beta_0$ and $\beta_1$ and to generate the data.}
\label{fig:dgp_two_steps}
\end{center}
\end{figure}

\noindent {\bf A two-step generation process.} Our requirements are to generate data from the distribution \eqref{eq:our_synthetic_model}, while controlling the difference of ARR between subgroup $G=0$ and subgroup $G=1$. More precisely, given a baseline hazard $h_0$, a law of $\bf X$, a prognostic factor $\gamma$ and a subgroup definition $G$, we would like to generate a continuum of heterogeneity levels. In other words we want to choose the coefficients $\beta_0$ and $\beta_1$ to achieve the levels of ARR desired. We denote by $\ARR_0(t)$ (resp. $\ARR_1(t)$) the ARR at time point $t$ in subgroup $G=0$ (resp. $G=1$). Unfortunately, the functions $\beta_0 \mapsto \ARR_0(1)$ and $\beta_1 \mapsto \ARR_1(1)$ have complex expressions and are not straightforward to compute, which leads us to the following two-step scheme.
\begin{enumerate}
    \item All other parameters being fixed, compute a Monte-Carlo estimate of the functions $\beta_0 \mapsto \ARR_0(t=1)$ and $\beta_1 \mapsto \ARR_1(t=1)$. 
    \item At time of generation, for fixed ARR levels to be reached in subgroups $G=0$ and $G=1$, select the corresponding parameters $\beta_0$ and $\beta_1$ thanks to step $1$. We reduce the scope and only focus on overall non-significant trials, meaning that $\ARR(t=1)=0$. This translates into a straightforward relation between $\beta_0$ and $\beta_1$ (see \eqref{eq:ARR_relation} in the supplementary material), and allows us to specify the heterogeneity level of an experiment by only making $\beta_1$ vary and choosing $\beta_0$ accordingly.
\end{enumerate}
We provide details on these two points in the supplementary material (Appendix \ref{appendix:dgp}). We also refer the reader to \cref{fig:dgp_two_steps} for an illustration.

\medskip

\noindent {\bf Our scenarios choices.} We take $\mathbf{X}$ to be a centered, isotropic Gaussian variable in dimension $p$ and explore three values for the dimension $p$: 20, 100 and 1000. The prognostic vectors $\gamma$ are provided in the supplementary material (see Section \ref{appendix:dgp}). All of our subgroup functions are of the form 
\begin{equation}\label{eq: subgroup_defn}
G(x) = \mathbf{1}_{x_i \geq -1, x_j \geq -1, x_k \geq -1, x_l \geq -1 }, 
\end{equation}
for some indexes $i, j, k, l$, whose choices allow us to vary the number of biomarkers that are both predictive and prognostic. The threshold value $-1$ is set so that the two subgroups have approximately equal proportions. Regarding censoring, we explore four scenarios: one without, and three with increasing censoring rates -- see Section \ref{appendix:dgp} of the supplementary material for more details on censoring, subgroups, sample size and other scenarios characteristics.

\medskip

\noindent {\bf Semi-synthetic generation.} We also consider a more realistic scenario for the law of the covariates $\mathbf{X}$ by relying on a real dataset. We use the HNSC cohort from the cancer genome atlas (TCGA), from which we extracted the RNAseq expression matrix, kept the first thousand most variable genes and applied normalization. The resulting matrix is of shape $503 \times 1000$. To emulate the law of $\mathbf{X}$, we take the empirical measure associated to the normalized expression matrix: a patient's descriptive variable $\mathbf{X}$ is a uniformly chosen row of the matrix. The law of $T$ given $\mathbf{X}$ is then provided by \eqref{eq:our_synthetic_model}. The prognostic vector $\gamma$ has been set by fitting a regularized multivariate Cox model predicting the overall survival from RNAseq expression. We define a subgroup function based on 5 variables that are not prognostic (meaning that their $\gamma$ coefficient is zero) and choose the thresholds for the 5 variables so that the subgroups of good and bad responders are balanced (respectively 248 and 255 patients). 

\medskip

\noindent {\bf The null and the alternative hypotheses.} In the following, the null hypothesis $H_0$ corresponds to the case $\beta_0 = \beta_1 = 0$, where no heterogeneity exists. The alternative hypotheses can be indexed by the value of the ARR in subgroup $G=1$: increasing values of this ARR correspond to stronger heterogeneity signal between $G=0$ and $G=1$.

\section{Research questions and associated metrics}
\label{sec:metrics}

Evaluating subgroup analysis methods requires to define relevant metrics. As described by \cite{sun2022}, there are 3 main questions to consider when studying heterogeneity of treatment effect. We investigate each of these questions under the null hypothesis (absence of heterogeneity) and under a spectrum of alternative hypotheses that have been previously described, with the help of our data generation process introduced in Section \ref{sec:dgp}. 
\begin{enumerate}
    \item \label{question:1} {\bf Existence of heterogeneity of treatment effect in the population.} The existence of heterogeneity of treatment effect relates to assessing whether there is significant evidence against a model of homogeneous treatment effect. One usually relies on an appropriate statistical test where the null hypothesis is the absence of heterogeneity. We report the type I error rate and the power of each method.
    
    \item \label{question:2} {\bf Identification of variables responsible for the heterogeneity.} Establishing the variables responsible for the heterogeneity of treatment effect corresponds to identifying the variables that are predictive of the treatment effect, possibly providing a ranking of variables from most to least predictive of the heterogeneity. Each method outputs a list of the variables deemed important regarding heterogeneity. Under the null hypothesis, we expect that all variables have the same probability of being top-ranked as predictive. When there is heterogeneity, the variables ranked as most predictive by the method under study should be compared to the set of variables defining the subgroup of good responders in the ground truth. In this context, we use two metrics. The first one consists of computing the number of times the top-ranked variable of a model is indeed a predictive variable (that is, the probability of the top-ranked variable to be predictive). The second metric is more global and envisions the problem as a classification task where one wants to predict whether a variable is predictive or not: as such, we compute the area under the precision-recall curve (referred to as the averaged precision score in the rest of this paper).

    \item \label{question:3} {\bf Identification of a subgroup of good responders.} Identifying a subgroup of good responders corresponds to a binary classification problem where patients are either assigned to the subgroup of good responders or to the subgroup of bad responders. The ground truth being known, we assess the ability of each method to appropriately classify samples in the two subgroups, considering the method accuracy. 

\end{enumerate}

\section{Subgroup analysis methods}
\label{sec:methods}

\subsection{Selected methods}

\noindent We selected 9 methods out of the identified subgroup analysis methods. Our selection was informed by the similarities between the methods, the existing implementations, and the types of outcomes handled by such methods -- see \cref{table:methods1} in supplementary material for more detailed information. We implemented them all in Python. The rest of this section is dedicated to giving a short description of each method. Recall that we denote by $\mathbf{X} = \{X_{1}, ..., X_{p}\}$ the covariates, $W$ the treatment variable, and $T$ the time-to-event variable. Because of how diverse subgroup analysis methods are, they differ in terms of output and thus of how each answers the research questions described in Section \ref{sec:metrics} -- see Sections \ref{method_rq1}, \ref{method_rq2}, \ref{method_rq3}. 

%for more information and Appendix A.1 for a discussion of the specifics of each method. 
%Indeed, adaptation to the time-to-event framework is far from being straightforward and several existing methods -- such as OWL \citep{zhao2012} or STIMA \citep{dusseldorp2010} -- cannot be adapted to it. 

\medskip

\noindent{\textbf{Univariate methods (Interaction and t-test).}} The interaction test is probably one of the most canonical test performed in biostatistics. For each covariate $X_{j}$, it consists of fitting a Cox model predicting the outcome $T$ from $X_{j}$ and the interaction term $X_{j}W$. The nullity of the coefficient of the interaction term in the fitted model is then tested; this method corresponds to the Univariate Interaction method. We also include the Univariate t-test method in our analysis, where for each covariate $X_{j}$, the data is split into two parts according to the median value of $X_{j}$, and a difference of treatment effect between the subgroups is tested.

\medskip

\noindent{\textbf{Multivariate methods (Cox and Tree).}} Multivariate methods have been introduced by \cite{chernozhukov2018} for subgroup analysis, but are more generally based on estimating a CATE. In our case, the absolute risk reduction (ARR) corresponds to the CATE and we estimate it with what is called a meta-learner \citep{kunzel2019metalearners}. We restrict our investigations to an S-learner, meaning that a ML-model $\hat{S}(t,\mathbf{X}, W)$ is trained to infer the survival function from $\mathbf{X}$ and $W$ at time $t$, and then estimate the ARR as the difference $\hat{S}(t, \mathbf{X}, 1) - \hat{S}(t, \mathbf{X}, 0)$. This corresponds to the benefit at time $t$ of receiving the treatment for an individual with characteristics $\mathbf{X}$. In practice, we split the data into training and test sets, estimate the ARR on the train set and then predict a subgroup of good responders on the test set by thresholding the estimated ARR: we set the subgroup of good responders to correspond to samples with $ARR \geq 0$. We decided to use two types of models to infer the survival function of the form $h(t \, | \, \mathbf{X}, W) = h_0(t) \exp\left( \beta_0 W + \beta_i \mathbf{X} + \beta_j W \mathbf{X} \right)$: a multivariate Cox model and a tree-based model. In view of our data generating process \eqref{eq:our_synthetic_model}, the latter is well-specified while the former is not. Hyperparameter tuning could be performed for the multivariate tree model but is not performed in this benchmark as it is not a goal of our work. 

\medskip

\noindent{\textbf{Model-based partitioning (MOB).}} This method was initially introduced for prediction by \cite{zeileis2008model}, and then adapted for subgroup analysis by \cite{seibold2016}. It is a tree-based algorithm consisting of fitting, at each node, a Cox model predicting the outcome using only the treatment variable $W$, and then identifying a variable to split on based on the correlations between the model score residuals and each covariate. The selected variable is the one corresponding to the smallest Bonferroni-adjusted p-value of correlation with the model residuals, and the splitting value is selected among its quantiles. The underlying rationale is that high correlation corresponds to the existence of an interaction between the variable and the treatment.

\medskip

\noindent{\textbf{Interaction tree (ITree).}} It is a tree-based method introduced by \cite{su2008}. At each node, we fit for each covariate $X_{j}$ a Cox model including as covariates the treatment variable $W$, a thresholded version of $X_{j}$ $Z_j:=\mathbf{1}_{X_j \leq c}$, and an interaction term $Z_jW$. Several values of thresholds, usually at several quantiles of $X_j$, are tested. The variable to split on is selected as the variable leading to the smallest Bonferroni-adjusted p-value when testing for the nullity of the interaction term coefficient.

\medskip

\noindent{\textbf{Subgroup identification based on differential effect search (SIDES).}} This method, introduced by \cite{lipkovich2011}, is based on recursively partitioning the observations into two groups such that the treatment effect is maximized in one subgroup compared to the other. At each split, the covariate space is searched looking for the covariates inducing subgroups of maximized treatment effect difference; multiple subgroups of interest (i.e. subgroups of good responders) are identified. The SIDES method is tree-like but as opposed to classic tree-based methods, child nodes at each level are not complementary (i.e. they are not created by a split on the same variable).
%and as such there can be more than 2 child nodes. 

\medskip

\noindent{\textbf{Sequential batting (SeqBT).}} The method was introduced by \cite{huang2017}. It consists of identifying a subgroup with maximized treatment effect by defining a multiplicative rule of the form $i(X)=\prod_i \mathbf{1}_{X_i \leq c_i}$ to classify patients as good or bad responders. The multiplicative index $i(X)$ is iteratively defined. At each iteration, several new binary rules are tested: the quality of the updated $i(X)$ is measured with an interaction test, by fitting a Cox model with covariates $W$, $i(X)$ and $i(X)W$. The best $i(X)$ update is considered to be the one leading to the smallest p-value when testing for the nullity of the interaction coefficient. 
%We define the final classification rule by fitting the sequential batting algorithm on training samples. 

\medskip

\noindent{\textbf{Adaptative Refinement by Directed Peeling (ARDP).}} This method, introduced by \cite{leblanc2005} and adapted for predictive covariates by \cite{patel2016}, is a greedy method that iteratively constructs a subgroup of good responders. At each step, the subgroup is updated by removing samples that are outliers with respect to a given covariate, which is the one along which the peeling leads to the largest treatment effect at this step.
By fitting the ARDP algorithm on training samples, the iterative construction of the subgroup allows one to obtain a rule, refined at each step, defining the good responders. 

\subsection{About hyperparameter tuning}

Many methods presented in this benchmark are tree-like methods (Multivariate Tree, ITree, MOB and SIDES) and as such involve several hyperparameters that can be fine-tuned. While parameter tuning is standard, it is not the main focus of this benchmark: we only aim at providing a high level comparison of selected methods and fix some of their hyperparameters as done by \cite{sun2022}, for instance. In particular, for tree-like methods, the number of splits considered for each variable at each node is fixed, as well as the minimal acceptable size of a child node. The depth of the SIDES tree-like structure is also fixed, as well as the percentage of samples peeled at each step for the ARDP algorithm.
%for SIDES, ITree and MOB: the number of splits considered for each variable at each node is set to 3 and the minimal acceptable size of a child node is equal to 10\% of total sample size. This minimal acceptable size is set to 5\% for the Multivariate Tree method. For SIDES, we also enforce a subgroup of maximum 4 covariates (depth of the tree-like structure). Finally for ARDP, 30\% of samples are peeled at each step and up until the subgroup reaches a number of samples equal to 10\% the total size. 
Similarly to hyperparameter tuning, regularization tuning is also not in scope of this work and is fixed to $0.1$. 

\subsection{How to answer each research question}

Each method has its specifics to answer the three research questions described in Section 
\ref{sec:metrics}. We rely on two datasets, each generated by our data generation process (Section \ref{sec:dgp}):
\begin{enumerate}
    \item {\bf A discovery dataset.} This is the dataset that is {\it observed} by the practitioner. For methods that are primarily designed to answer research question \ref{question:3} (Multivariate methods, SIDES, SeqBT and ARDP), it is further splitted into two parts:
    \vspace{-0.2cm}
    \begin{enumerate}
        \item {\bf A training set.} Subset of the discovery dataset used to fit methods.
        \item {\bf A testing set.} Subpart of the discovery dataset used to compute an unbiased p-value for the first research question, following \cite{chernozhukov2018}.
    \end{enumerate}
    \item {\bf A validation dataset.} This dataset is only used for evaluating research question \ref{question:3} i.e. the generalization capabilities of the methods.
\end{enumerate}

\subsubsection{Answering research question 1: Existence of heterogeneity}
\label{method_rq1}

There are two ways to answer the research question \ref{question:1}.
For univariate methods, as well as MOB and ITree, the existence of heterogeneity is assessed directly during the model fitting step. The smallest Bonferroni-adjusted p-value among all univariate tests is reported as measure of heterogeneity existence; for MOB and ITree the Bonferroni-adjusted p-value of the root node split is reported. 
For the multivariate methods, SIDES, SeqBT and ARDP, which are predictive methods (see above), the existence of heterogeneity is assessed after the model fit and after the subgroup of good responders (research question \ref{question:3}) is defined, by testing for the treatment effect difference between the two subgroups obtained; the corresponding p-value is reported.

\subsubsection{Answering research question 2: Identification of predictive variables}
\label{method_rq2}

There are three ways to rank variables in terms of importance (and thus to obtain the top-ranked covariate in terms of induced heterogeneity). 
First, rankings of univariate and multivariate Cox methods (and thus top-ranked covariate identification) are computed by ranking the p-values of each covariate-treatment interaction term obtained from model fitting: the smallest the p-value, the highest the covariate position in the ranking. Of note, the multivariate tree method doesn't provide a way to rank variables.
Then, SIDES, SeqBT and ARDP rankings are rule-based rankings: once the good responders subgroup rule is defined following research question \ref{question:3}, the ranking is produced based on the covariates used in the subgroup rule. Variables are ranked by the inverse of their appearance in subgroup definition -- the last variable in the rule corresponds to the variable used last to partition patients and is of least importance.
For MOB and ITree, we follow \cite{sun2022} who defined the top-ranked variable as the variable used to define the first split of the tree. To produce the rankings of these two tree-based methods we generalize this top variable measure by defining a new feature importance measure, presented below.

\medskip

\noindent{\textbf{A new feature importance formula.}} Building on existing feature importance formula for tree-supervised methods (see for instance \cite{scornet2023trees}), we define the importance of a $X_i$ as:
\[     I(X_i) := \sum_{v \in V(X_i)} \frac{1}{p_{v}}  \cdot \frac{S_{v}}{S} , \] 
where $V(X_i)$ is the set of nodes using the variable $X_i$ as split, $S$ is the total sample size, $S_{v}$ the size of node $v$ and $p_{v}$ the p-value of the test used to select $X_i$ as the split variable at node $v$. In words, a variable has inflated importance when a node at the beginning of the tree uses it and when its associated $p$-value is small. We think this intuitive formula could be of independent interest for future research.

\subsubsection{Answering research question 3: Identification of good responders}
\label{method_rq3}

All methods can be used for identification of good responders, with the exception of univariate methods (for which we define a proxy below). However, methods differ in terms of how they are fitted for this task.
For multivariate methods, because we are interested in non-significant clinical trials where the subgroups have opposite treatment effects, we threshold the treatment effect function at 0 to separate samples of positive treatment effect (good responders) from the ones of negative treatment effect (bad responders). 
Tree-based (MOB, ITree) and tree-like (SIDES) methods allow the identification of the final subgroup (tree leaf) that has the highest treatment effect. The classification rule of this leaf defines the subgroup of good responders.
For rule-based methods (SeqBT, ARDP), the final rule is used to define the subgroup of good responders.
Finally, as means of comparison of univariate methods to other methods, we propose a simple threshold at the top-ranked variable median value to constitute good and bad responders subgroups.

\section{Results}
\label{sec:results}

This section is dedicated to the presentation and analysis of our experiments. Because we consider a wide range of scenarios,  for ease of reading and to convey the main findings, we focus on a subset of the experiments. Additional results may be found in the supplementary material -- see Appendix \ref{app:benchmark}.

\medskip

\noindent {\bf Continuum of heterogeneity levels.} In the following figures, the ARR value on the x-axis corresponds to the ARR level achieved in the subgroup of good responders. As already discussed in Section \ref{sec:dgp}, the associated ARR value in the subgroup of bad responders is computed so that the trial is overall non-significant. Since our subgroup definitions are such that good and bad responders subgroups are approximately balanced, the ARR value in the subgroup of bad responders is roughly equal to the opposite value of the ARR in the subgroup of good responders. This results in a difference of ARR between good and bad responders being equal to twice the value of the x-axis. The plots are always drawn for the non-censored scenario. This is because the censoring rate not only depends on the censoring scenario, but also on the ARR point we consider. As a result, similar plots for the three censoring scenarios would not compare similar censoring rates. The interested reader can find performances of these censored scenarios in several tables in the supplementary material (Appendix \ref{app:rqs_table}). In particular, performance metrics degrade as censoring increases, as expected.

\medskip

\noindent {\bf Complexity considerations.} Several methods of our benchmark were initially designed for small dimensional settings. We conducted computational complexity analysis and found that the run times of SIDES and SeqBT were of higher orders of magnitude (see \cref{fig:complexity} in the supplementary material) compared to the other methods. As a result, we did not investigate their performance for $p=100$ and $p=1000$.

\medskip

\noindent {\bf Confidence intervals.} To ease reading of the plots, we did not include confidence intervals. Nevertheless, the interested reader may find the 95\% intervals reported in the tables of the supplementary material (Appendix \ref{app:rqs_table}).

%describe "abbreviations" used for each method names here again? in particular multivariate and univariate submethods?

%\noindent {\bf Plots of the main paper}

%\smallskip

%\noindent As described in the paragraph ``our scenarios choices" in section \ref{sec:dgp}, there are lots of scenarios. We restrict the number of results shared in this main paper, and release other plots in the supplementary material. In particular, except explicitly stated otherwise, we will always consider the setting where the sampling size is $500$, the train/test split is $0.5/0.5$, $0$, the scaling rate is $0.05$ and the number of predictive variables that are prognostic is $0$. This is motivated by the fact that neither the sampling size, nor the train/test split or the scaling rate are affecting the ranking of the methods. We illustrate this for the scaling parameter in Figure [REF]. Similar figures for the sample size and train/test split can be found in the supplementary material.

%\begin{figure}[htb]
%\begin{center}
%\includegraphics[scale=0.4]{figures/scale_move.pdf}
%\caption{Censorship scale moves. $p=20$, prog0, train 0.5, size 500.}
%\end{center}
%\end{figure}

\subsection{ Research question 1 -- Existence of heterogeneity of treatment effect}

\smallskip

\noindent {\bf Type I error.} To assess the methods' ability to capture the existence of treatment effect heterogeneity, we first report the type I error of each method -- see Table \ref{table:type1err}. In particular, the type I error is evaluated in 3 different data generation scenarios of increasing dimension (20, 100, 1000), over 1000 repetitions. 
Of note, it is first worth mentioning that for some methods (Multivariate Tree and ARDP), the type I error rate is biased towards 0.05 exactly because of an implementation choice that consists of returning a random p-value between 0 and 1 when all samples are predicted to belong to the same subgroup. It is appropriate since predicting all samples in the same group corresponds to a situation of treatment effect homogeneity. Looking at Table \ref{table:type1err}, univariate interaction and interaction tree methods are notably overconservative and this conservative character worsens with increasing dimension. 
SIDES does not control the type I error well; it is expected from our implementation. Indeed as discussed above, the SIDES algorithm is highly computationally expensive even in small (p=20) dimension, in particular because of the repeated permutation setup that is implemented to control the type I error; as such, this step was removed from our implementation, which explains the uncontrolled type I error.
As a mean of comparison this method is still included here but computation-wise we believe that this method is well suited for even smaller dimension (p$\le$5) as implied by \cite{lipkovich2011}. 

\begin{table}[htb]
\begin{center}
\caption{Type I error of each method in dimensions $p=20, 100$ and $1000$. Subgroup definitions use four predictive variables that are not prognostic. We repeat the DGP 1000 times for the ARR=0 point, with sampling size 500, train/test split $0.5/0.5$ and under the first censoring scenario.}
\label{table:type1err}
\begin{tabular}{lccc}
\hline
 Scenario & p=20 & p=100 & p=1000  \\
%Description 1 & Description 2 & Description 3\\
\hline
 Univariate interaction & 0.028 & 0.007 & 0.001 \\
 Univariate t-test      & 0.035 & 0.052 & 0.052 \\
 Multivariate Cox       & 0.042 & 0.033 & 0.037 \\
 Multivariate Tree      & 0.051 & 0.051 & 0.051   \\
 MOB                    & 0.060 & 0.043 & 0.056  \\
 IT                     & 0.024 & 0.011 & 0.002  \\
 ARDP                   & 0.036 & 0.049 & 0.043 \\
 SIDES                  & 0.125 & --    & -- \\
 SeqBT                  & 0.042 & --    & -- \\
\hline
\end{tabular}
\end{center}
\end{table}

\medskip

\noindent {\bf Power analysis.} We perform a power analysis to pursue our investigation of the capacities of methods to capture heterogeneity of treatment effect. Methods are compared over a range of heterogeneity scenarios (from homogeneous treatment effect to high heterogeneity).
We include the oracle power curve for comparison, which consists of fitting a univariate model on the true subgroup indicator variable. This method constitutes the best subgroup identification method and thus represents the best possible performance, as it uses the ground truths directly.
At high heterogeneity, a lot of methods have similar power (univariate interaction, MOB and ITree, followed by SeqBT and multivariate tree). However, it is worth noticing the superiority of the univariate interaction, MOB and ITree methods, which surpass the others in low and intermediate heterogeneity settings.
Power decreases with dimension, but it does not affect how methods compare to each other.

\begin{figure}[htb]
\begin{center}
\includegraphics[scale=0.45]{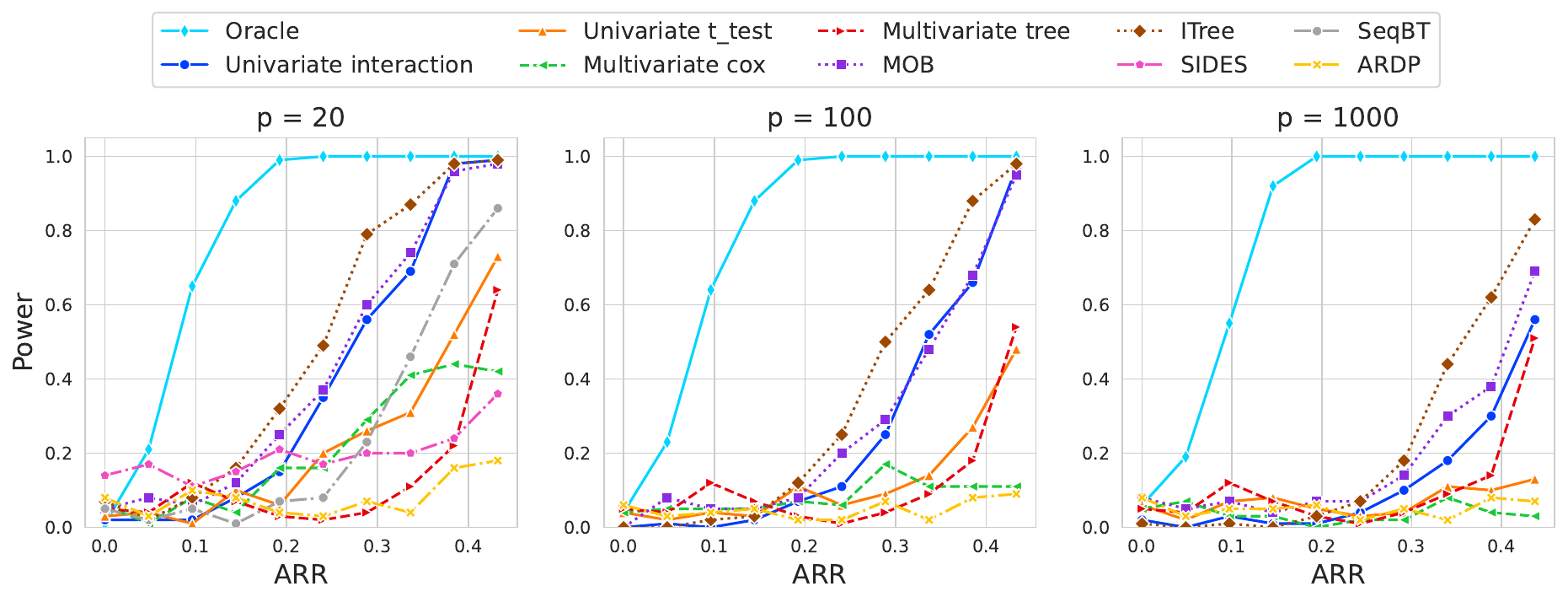}
\caption{Power analysis of each method in dimensions $p=20, 100$ and $1000$. Subgroup definitions use four predictive variables that are not prognostic. We repeat the DGP 100 times for 10 ARR points, with sampling size 500, train/test split $0.5/0.5$ and without censoring.}
\label{fig:power}
\end{center}
\end{figure}

\medskip

\begin{figure}
        \centering
        \includegraphics[width=0.92\linewidth]{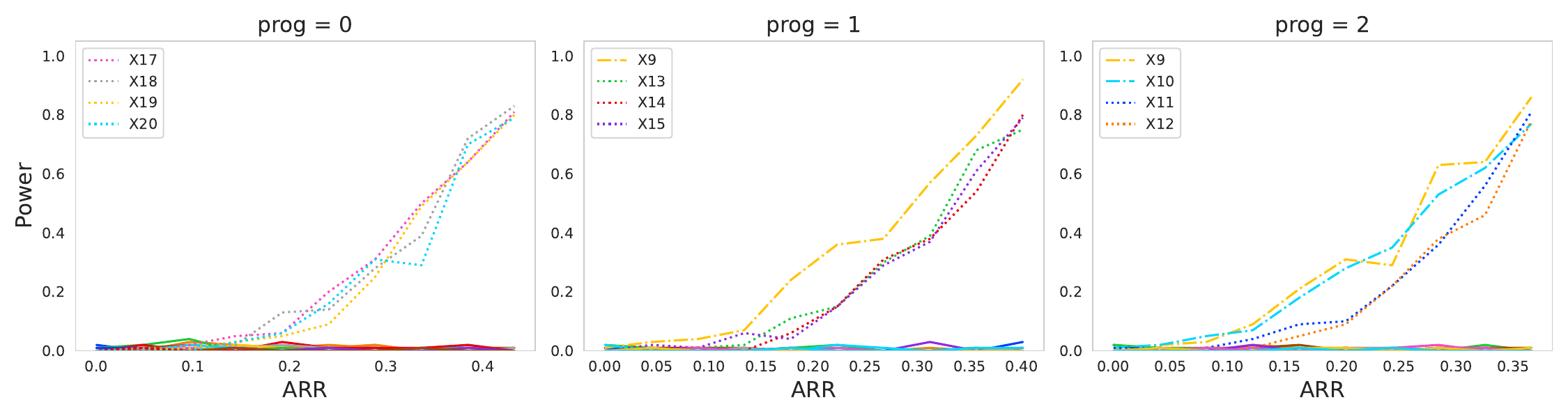}

\vspace{0.2cm}

        \centering
        \includegraphics[width=0.63\linewidth]{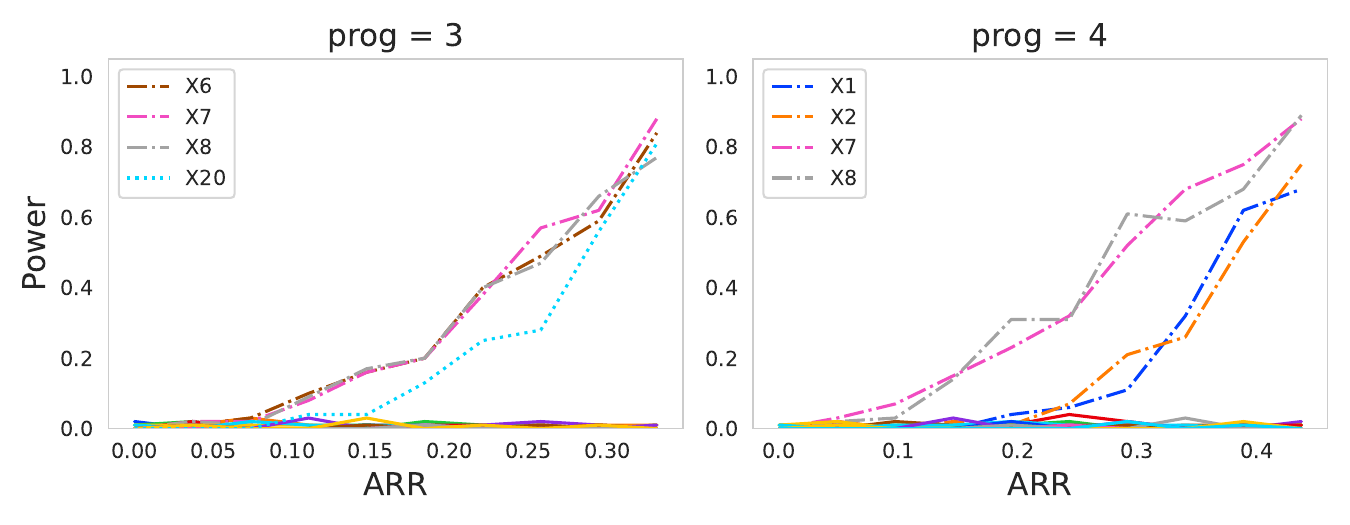}

    \caption{Power analysis of each univariate interaction test in dimensions $p=20$. We use a subgroup definition using 4 predictive variables and vary the number of prognostic variables among them, resulting in 5 scenarios. Power curves of variables that are predictive and prognostic are dash-dotted; dotted curves correspond to variables that are only predictive. We repet 100 times the DGP for 10 ARR points, with sampling size 500, train/test split $0.5/0.5$ and without censoring.}
\label{fig:zoomuniv}
\end{figure}

\noindent {\bf Zoom on the univariate interaction method -- Subgroup complexity.} We focus on the univariate interaction method since it appears to be one of the best methods to detect heterogeneity. In particular, we focus on univariate power curves in small dimension (p=20) for 5 different subgroups definition, where we vary the number of prognostic variables among the predictive ones -- see Appendix \ref{appendix:dgp} for precise definitions. The ability of the univariate interaction method to capture the existence of heterogeneity is due to the performance of the univariate models for each of the predictive variables, as shown by the dotted and dash-dotted power curves that clearly separate from the plain ones. Moreover, we see that with the exception of extreme heterogeneity settings, the power of univariate interaction method is dependent on the prognostic character of predictive variables: situations with subgroups defined with 1, 2, or 3 prognostic predictive variables show that prognostic information leads to increased power. Finally, more than the prognostic status of the predictive variable, it is the positive or negative prognostic aspect that influences the method's ability to detect heterogeneity, as observed in \cref{fig:zoomuniv}, bottom right: a univariate model built on a negative prognostic covariate has higher power compared to a univariate model built on a positive prognostic covariate.

%The horizontal axis corresponds to the ARR level in the subgroup of good responders, which by definition correlates to the heterogeneity intensity level

\subsection{Research question 2 -- Retrieval of predictive variables}

\smallskip

\noindent {\bf Probability that the top-ranked variable is predictive.} To evaluate if methods are able to retrieve the predictive variables (i.e. variables defining the subgroups), we analyze their respective rankings of variables. We first focus on the probability that the top-ranked variable is predictive. This allows us to perform a first assessment of the ability of methods to correctly identify variables responsible for the heterogeneity of treatment effect. 
We remind that the multivariate tree method does not provide a ranking of variables and is thus excluded from the research question \ref{question:2}. 
Univariate interaction, multivariate Cox, ITree and MOB methods appear superior to other methods when it comes to selecting a predictive variable as top-ranked, followed by SeqBT. 
Additionally, it is worth mentioning two phenomena happening in small dimension (p=20). First, when the heterogeneity of treatment effect is very large, the ability of the univariate t-test method to select a predictive top-ranked variable starts to drop. Secondly, in case of homogeneous treatment effect between subgroups (when ARR=0), SeqBT and multivariate Cox methods are biased -- this phenomenon is more visible when $p=20$ because of the scale of our plots. Indeed, in absence of heterogeneity, each variable should be selected as top-ranked equally, leading to a probability of selecting one of the four predictive variables as top-ranked equal to $4/p$.  

\begin{figure}[htb]
\begin{center}
\includegraphics[scale=0.43]{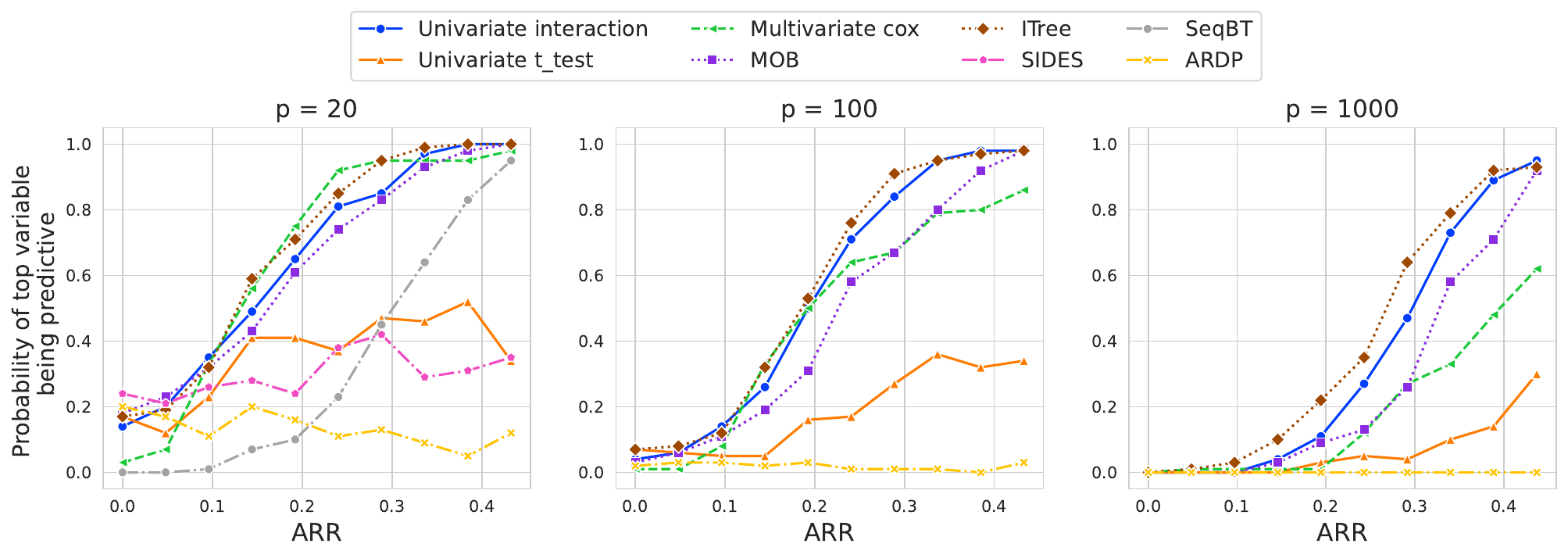}
\caption{Probability of the top-ranked variable to be predictive for each method in dimensions $p=20, 100$ and $1000$. Subgroup definitions use four predictive variables that are not prognostic. We repeat the DGP 100 times for 10 ARR points, with sampling size 500, train/test split $0.5/0.5$ and without censoring.}
\label{fig:topvar}
\end{center}
\end{figure}

\smallskip

\noindent {\bf Averaged precision score.} Using our newly defined feature importance measure, we further assess the ability of each method to correctly rank variables. We report the averaged precision scores across heterogeneity scenarios. Results are consistent with the top-ranked variable results, with an increased performance for the univariate interaction and multivariate Cox methods which score higher than other methods. This demonstrates that more than accurately picking a predictive variable as top-ranked, these methods are able to appropriately retrieve all variables responsible for the heterogeneity, with increased performance in large heterogeneity settings. 
We also still observe a drop of performance of the univariate t-test method in high heterogeneity scenarios.

\begin{figure}[htb]
\begin{center}
\includegraphics[scale=0.43]{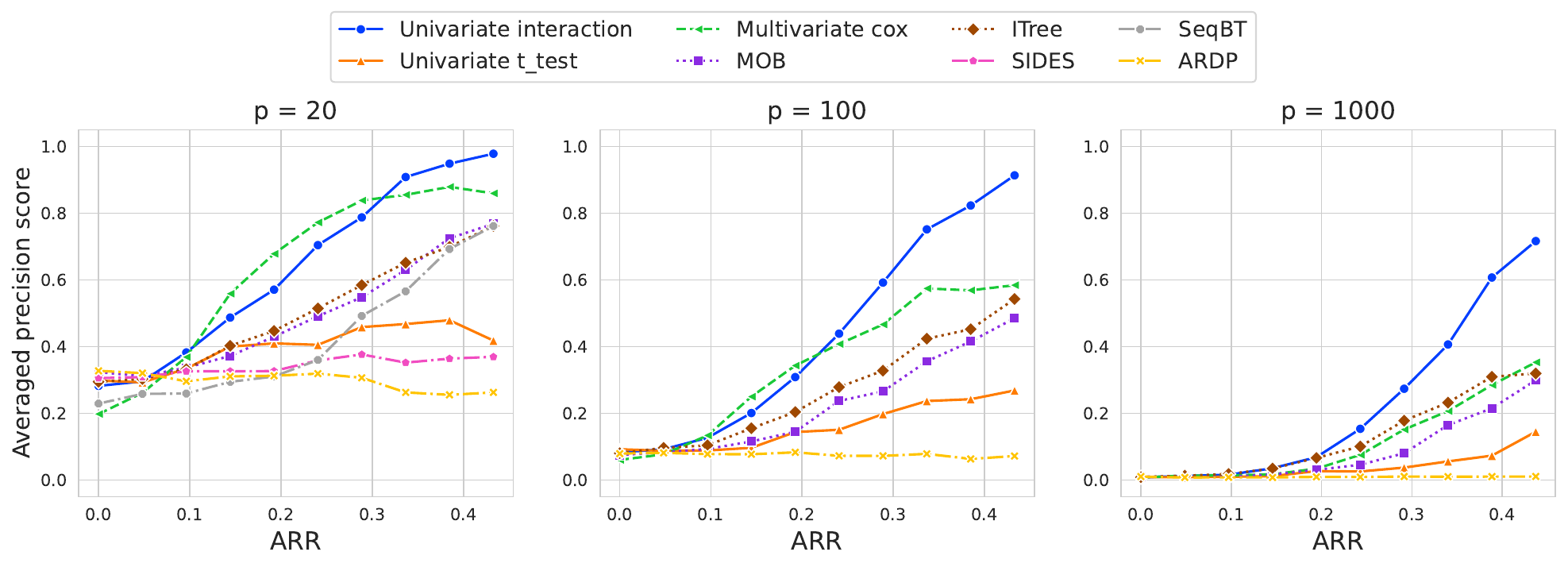}
\caption{Averaged precision scores of each methods in dimensions $p=20, 100$ and $1000$. Subgroup definitions use four predictive variables that are not prognostic. We repeat the DGP 100 times for 10 ARR points, with sampling size 500, train/test split $0.5/0.5$ and without censoring.}
\label{fig:precision}
\end{center}
\end{figure}

\begin{figure}[htb]
        \centering
        \includegraphics[width=0.8\linewidth, height=5.12cm]{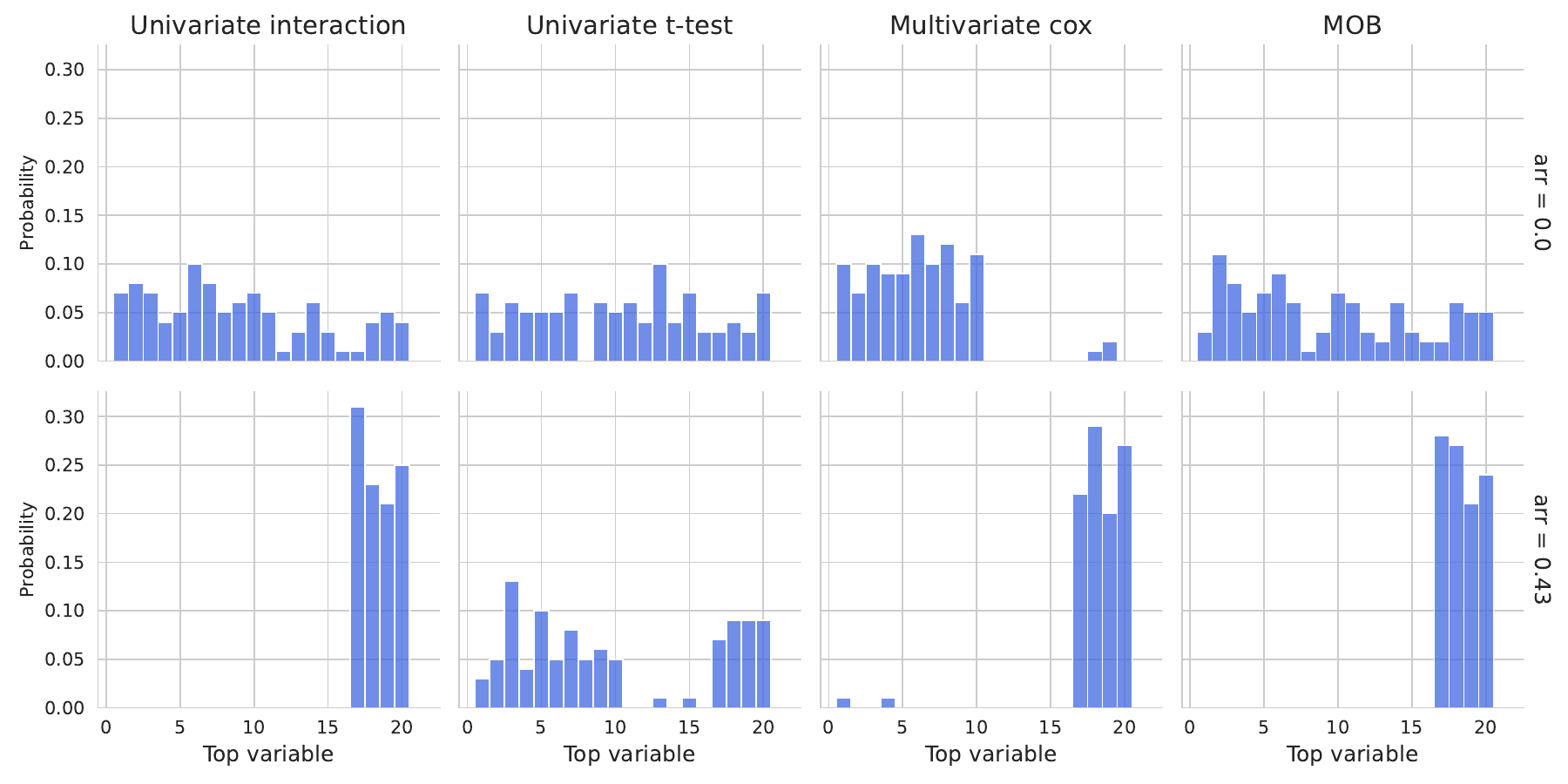}

        \centering
        \includegraphics[width=0.8\linewidth, height=5.12cm]{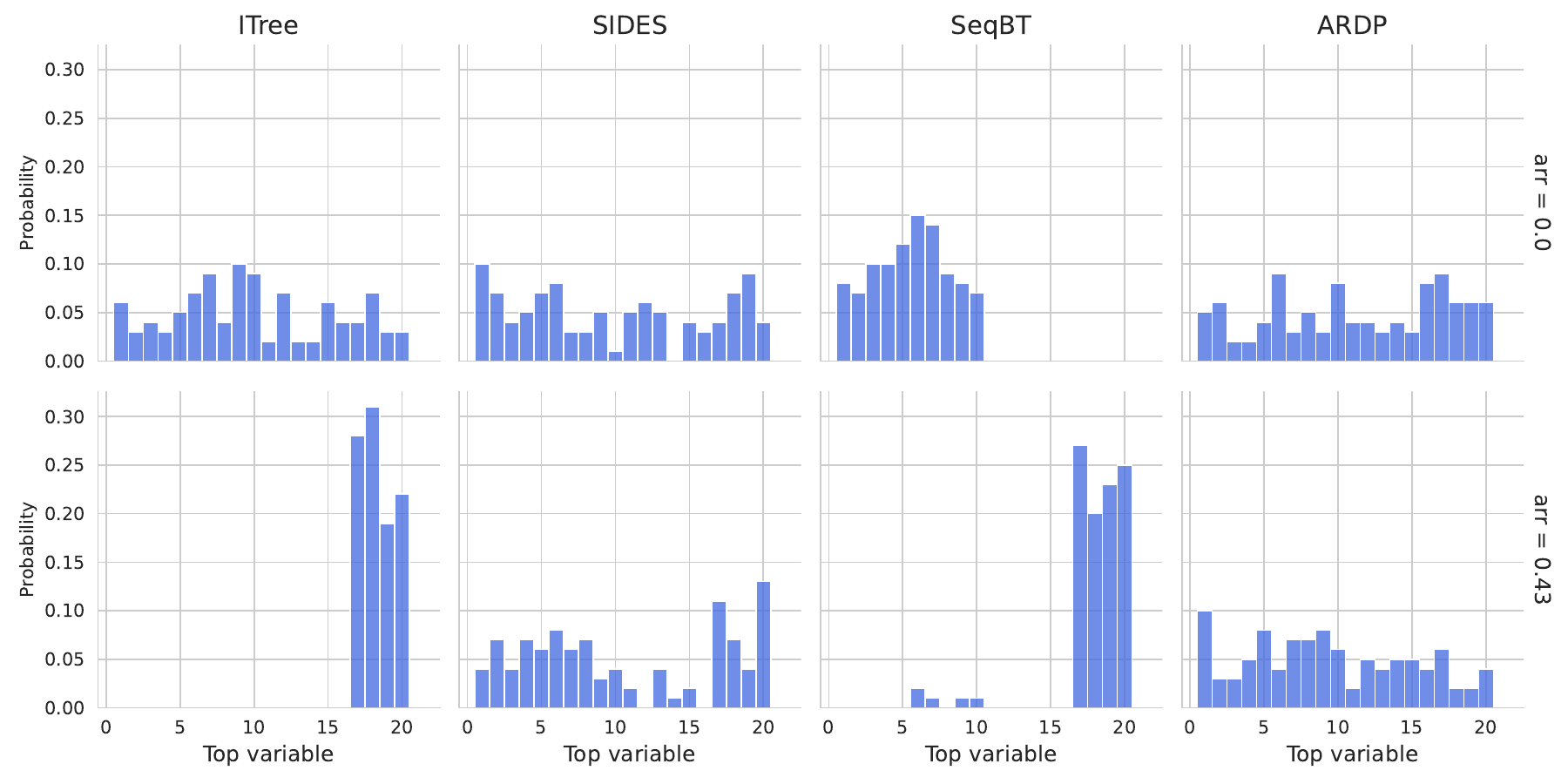}

    \caption{Histograms of the probabilities that each variable is top-ranked by the different methods in dimension $p=20$, in absence of heterogeneity and in high heterogeneity setting. Subgroup definition use four predictive variables that are not prognostic. More precisely, $X_1, \ldots, X_{10}$ are prognostic and $X_{17}, X_{18}, X_{19}, X_{20}$ are poredictive. We repeat the DGP 100 times for 10 ARR points, with sampling size 500, train/test split $0.5/0.5$ and without censoring.}
\end{figure}

\noindent {\bf Zoom on top-ranked variables in extreme heterogeneity scenarios.} To better understand the performance results of the identification of predictive variables, in particular focusing on the specific cases of {\it(i)} the univariate t-test performance drop and {\it(ii)} the SeqBT and multivariate cox methods' bias mentioned above, we focus on the small dimension setting (p=20) and on two extreme heterogeneity scenarios: homogeneity of treatment effect and very high heterogeneity. For these two scenarios, we show how many times (out of 100 repeats) each of the 20 variables is selected as the top-ranked variable. 
When there is no heterogeneity of treatment effect, the chances of being picked out as top-ranked variable should be equally distributed among the 20 variables, and we expect all variables to be picked out approximately the same number of times. Looking at \cref{fig:precision}, the multivariate cox and SeqBT methods show an unequal distribution of top-ranked picks across the 20 variables, as opposed to other methods. This confirms that as mentioned above, these two methods are biased, but most importantly they are biased in favor of the prognostic variables -- see Section \ref{appendix:dgp} in supplementary material for the prognostic vector definition in dimension p=20. In the high heterogeneity setting, as discussed before, five methods accurately pick out predictive variables: univariate interaction, multivariate Cox, ITree, MOB, and SeqBT. Moreover, the plots allow us to explain the previously observed drop of performance in predictive variables retrieval for the univariate t-test method. This is caused by this method no longer top-ranking predictive variables, but rather prognostic variables when the treatment effect heterogeneity is high.

\subsection{Research question 3 -- Identification of good responders subgroups}

\smallskip

\begin{figure}[htb]
\begin{center}
\includegraphics[scale=0.45]{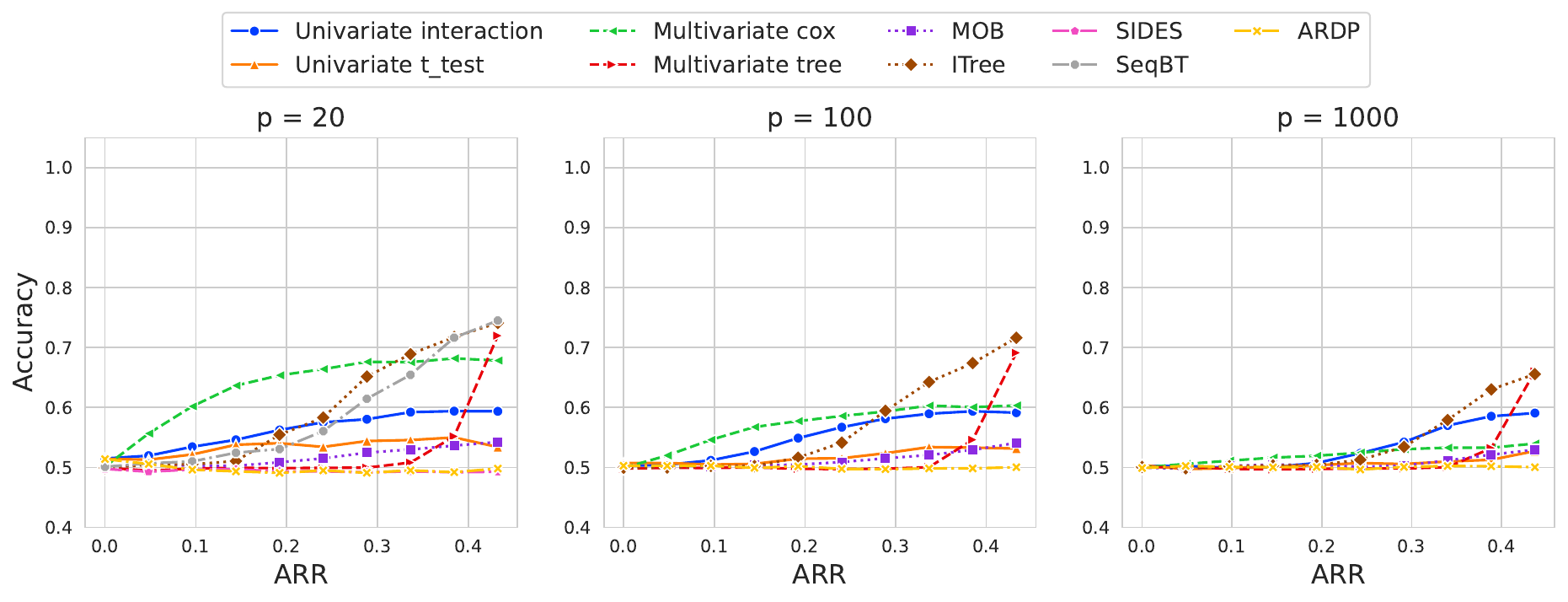}
\caption{Classification accuracy of each method in dimensions $p=20, 100$ and $1000$. Subgroup definitions use four predictive variables that are not prognostic. We repeat the DGP 100 times for 10 ARR points, with sampling size 500, train/test split $0.5/0.5$ and without censoring.}
\label{fig:accuracy}
\end{center}
\end{figure}

\noindent {\bf Classification accuracy.} Methods are then evaluated for their ability to identify the good responders, in other words for their performance in the binary classification task consisting of assigning samples to good and bad responders subgroups. We report the classification accuracy of each method in \cref{fig:accuracy}. 
Classification is a machine learning task; as such, machine learning approaches are in principle better suited to answer this research question than biostatistics approaches -- see \cref{sec:comparative}. Additionally, some methods do not provide straightforward ways to identify a subgroup, such as univariate methods which are not made for this purpose.
Four methods show high accuracy of subgroup prediction in high heterogeneity settings: the multivariate methods (Cox and tree), ITree, and SeqBT. Additionally the multivariate Cox method shows higher accuracy of subgroups prediction and differentiates from other methods in low heterogeneity setting. However, it reaches a plateau at high heterogeneity while three others clearly perform better. 
This is expected since these three methods are well-specified with respect to the data generation process and subgroup definitions, while the multivariate cox method is not.
Moreover, and as already discussed, we did not perform any hyperparameter tuning, which could further improve the performance of the multivariate tree and ITree methods (among others). 
Machine learning approaches thus tend to perform better than classic biostatistics methods to identify subgroups of good responders.

\subsection{Semi-synthetic experiment}

We performed an additional benchmarking experiment using semi-synthetic data, in dimension $p=1000$ and with a subgroup definition involving 5 predictive variables -- one more than in \eqref{eq: subgroup_defn} -- that are not prognostic. We repeat the DGP 100 times for 10 ARR points, with sampling size 500, train/test split $0.5/0.5$ and under the first censoring scenario.

\smallskip

\noindent {\bf Existence of heterogeneity.} In Table \ref{table:semi_synth_rq1}, we report the power of each method for each ARR point. Univariate interaction and ITree methods perform the best, while other methods exhibit limited power (univariate t-test, multivariate Cox, MOB), if not none at all (multivariate tree, ARDP). Notice that the univariate interaction method and ITree are very conservative regarding type I error.

\begin{table}[htb]
\begin{center}
\caption{Power analysis of each method for the semi-synthetic experiment.}
\label{table:semi_synth_rq1}
\begin{tabular}{c|cccccccc}
\hline
ARR & Oracle & U interaction & U t-test & M Cox & M Tree & MOB & ITree & ARDP \\
\hline
0.00 & 0.07 & 0.00 & 0.03 & 0.03 & 0.05 & 0.03 & 0.00 & 0.02 \\
0.04 & 0.09 & 0.00 & 0.04 & 0.03 & 0.04 & 0.04 & 0.00 & 0.07 \\
0.08 & 0.32 & 0.00 & 0.05 & 0.05 & 0.12 & 0.04 & 0.00 & 0.06 \\
0.13 & 0.64 & 0.04 & 0.08 & 0.06 & 0.07 & 0.06 & 0.04 & 0.04 \\
0.17 & 0.94 & 0.07 & 0.05 & 0.05 & 0.03 & 0.08 & 0.13 & 0.01 \\
0.21 & 1.00 & 0.23 & 0.09 & 0.04 & 0.01 & 0.14 & 0.24 & 0.08 \\
0.25 & 1.00 & 0.51 & 0.09 & 0.13 & 0.04 & 0.29 & 0.55 & 0.07 \\
0.29 & 1.00 & 0.80 & 0.12 & 0.22 & 0.08 & 0.35 & 0.90 & 0.09 \\
0.34 & 1.00 & 0.96 & 0.18 & 0.43 & 0.04 & 0.71 & 0.98 & 0.10 \\
0.38 & 1.00 & 1.00 & 0.42 & 0.69 & 0.12 & 0.95 & 1.00 & 0.18 \\
\hline
\end{tabular}
\end{center}
\end{table}

\noindent {\bf Predictive biomarkers identification.} In Table \ref{table:semi_synth_rq2}, we report the averaged precision scores for each method. Methods do not perform well, with the exception of ITree and the univariate interaction method which are slightly able to retrieve predictive biomarkers in high heterogeneity settings. The same phenomenon occurs if one looks at the probability that the top ranked variable is predictive. 

\begin{table}[htb]
\begin{center}
\caption{Averaged precision score of each method for the semi-synthetic experiment.}
\label{table:semi_synth_rq2}
\begin{tabular}{c|cccccc}
\hline
ARR & U interaction & U t-test & M Cox & MOB & ITree & ARDP \\
\hline
0.00 & 0.01 & 0.01 & 0.01 & 0.01 & 0.01 & 0.04 \\
0.04 & 0.01 & 0.01 & 0.01 & 0.01 & 0.02 & 0.04 \\
0.08 & 0.02 & 0.01 & 0.01 & 0.01 & 0.02 & 0.03 \\
0.13 & 0.04 & 0.02 & 0.01 & 0.02 & 0.02 & 0.04 \\
0.17 & 0.07 & 0.03 & 0.02 & 0.03 & 0.05 & 0.04 \\
0.21 & 0.09 & 0.03 & 0.01 & 0.03 & 0.04 & 0.04 \\
0.25 & 0.11 & 0.04 & 0.03 & 0.04 & 0.04 & 0.04 \\
0.29 & 0.13 & 0.04 & 0.03 & 0.05 & 0.07 & 0.05 \\
0.34 & 0.13 & 0.05 & 0.03 & 0.08 & 0.12 & 0.05 \\
0.38 & 0.14 & 0.04 & 0.05 & 0.11 & 0.17 & 0.04 \\
\hline
\end{tabular}
\end{center}
\end{table}

\noindent {\bf Good responders identification.} In Table \ref{table:semi_synth_rq3}, we report the accuracy of each method. In contrast with our synthetic experiments, the multivariate tree method does not exhibit good performance. As already discussed, the fact that we did not optimize its numerous hyperparameters could explain this phenomenon. Interestingly, multivariate Cox and ITree have good performance ($\sim70-80\%$) in the highest heterogeneity scenarios. This supports the fact that ML-like methods are able to detect good responders, even in high-dimensional settings. 

\begin{table}[htb]
\begin{center}
\caption{Accuracy of each method for the semi-synthetic experiment.}
\label{table:semi_synth_rq3}
\begin{tabular}{c|ccccccc}
\hline
ARR & U interaction & U t-test & M Cox & M Tree & MOB & ITree & ARDP \\
\hline
0.00 & 0.52 & 0.51 & 0.50 & 0.49 & 0.51 & 0.51 & 0.53 \\
0.04 & 0.52 & 0.52 & 0.52 & 0.49 & 0.51 & 0.51 & 0.54 \\
0.08 & 0.52 & 0.51 & 0.54 & 0.49 & 0.51 & 0.51 & 0.52 \\
0.13 & 0.54 & 0.52 & 0.57 & 0.49 & 0.51 & 0.51 & 0.53 \\
0.17 & 0.57 & 0.53 & 0.58 & 0.49 & 0.51 & 0.53 & 0.54 \\
0.21 & 0.57 & 0.53 & 0.59 & 0.49 & 0.51 & 0.57 & 0.55 \\
0.25 & 0.57 & 0.53 & 0.63 & 0.49 & 0.53 & 0.64 & 0.56 \\
0.29 & 0.59 & 0.53 & 0.63 & 0.49 & 0.54 & 0.73 & 0.56 \\
0.34 & 0.58 & 0.54 & 0.66 & 0.49 & 0.57 & 0.79 & 0.55 \\
0.38 & 0.58 & 0.54 & 0.69 & 0.52 & 0.57 & 0.82 & 0.56 \\
\hline
\end{tabular}
\end{center}
\end{table}

\bigskip

\section{Conclusion}
\label{sec:discussion}

In this paper we compared 9 subgroup analysis methods using both synthetic and semi-synthetic data, emulating time-to-event RCTs in a wide variety of heterogeneity scenarios. In particular, we introduced a new time-to-event data generating process allowing users to define complex subgroups of good and bad responders, and to precisely control the treatment effect heterogeneity level between the two subgroups. We also implemented multiple subgroup analysis methods some of which were not available in Python, and some not publicly available in any programming language. Each of these methods has been evaluated in light of three different research questions, following \cite{sun2022}. We hope our results, together with the benchmarking code that we are open sourcing in our \href{https://github.com/owkin/hte}{\texttt{hte}} package, will help further investigations and subsequent design of subgroup analysis in RCTs. 

%Despite the high prevalence of time-to-event endpoints in RCTs, most of existing benchmarks have been conducted on binary or continuous outcomes.

Not all methods are suited for all research questions. Univariate and multivariate Cox methods are in general performing better than other methods when it comes to heterogeneity identification, mostly by design (question \ref{question:1}) or predictive biomarkers identification (question \ref{question:2}), but ML methods are better suited for good responders identification (question \ref{question:3}). Note that the performances of these ML-based methods could be further improved by relying on hyper-parameter optimization, which was not considered in this benchmark paper as we mainly aimed at providing high level comparison of methods, as in previous benchmarks such as \cite{sun2022}. It is also worth noticing the consistently good performance of ITree, which as previously discussed can be classified as a biostatistics and ML-based approach. As a takeaway, our recommendation is to take the best of each set of methods: {\it (i)} use biostatistics methods to assess if heterogeneity is significant within a trial and which variables are responsible for it, and then {\it (ii)} try to identify the good responders with ML-based methods, if sufficient evidence supports existence of heterogeneity, as this potentially allows for identification of complex subgroups. For instance, one could first rely on univariate interaction tests, to select the most relevant variables and then train a multivariate Cox or ITree model to estimate individual treatment effects. 

Let us finally emphasize that the main focus of this paper is to benchmark the performance of subgroup analysis methods. In real life settings, subgroup analysis is always about the trade-off between different subgroup characteristics, most notably between the treatment effect size in the subgroup and its size -- the number of patients it encompasses. This can be handled in different ways depending on the method used: for multivariate methods the threshold used to split good responders from bad ones can be adapted to favor a higher treatment effect or a larger subgroup; for tree-based methods multiple leafs can be combined to define a subgroup. In particular, methods such as SIDES and ARDP that do not perform well in these benchmark settings are nonetheless of high interest for trade-off considerations as they produced nested subgroups of good responders, allowing one to take into account the effect size, the sample size, and the predictive variables to define subgroups of good responders.
Finally more than the trade-off between the effect size and the subgroup size, it is the actionability of a subgroup that is crucial in real-life settings. Subgroups are indeed of interest if they can be reproduced, found in the real world, and actually lead to significant findings. For instance, complex ML algorithms are able to fit highly complex heterogeneity patterns at the price of a loss of interpretability, making their findings difficult to be translated in the clinic. In such circumstances, post-hoc interpretation methods could be of interest to extract actionable inclusion criteria from complex decision rules: this endeavor has not been discussed in this paper but could be the subject of future research avenues.

\bigskip

\medskip

\noindent {\bf Acknowledgements.}
We would like to thank Paul Trichelair and Kevin Elgui for fruitful preliminary discussions about this work. We also thank Jean-Philippe Vert, Jean Ogier du Terrail and Geneviève Robin for their very insightful comments and suggestions which greatly improved the quality of the manuscript. Moreover, we would like to thank Omar Darwiche Domingues for his continuous support in developing the code base, as well as Thomas Chaigneau for his timely suggestions on the pipeline. 
\vspace*{1pc}

%\noindent {\bf{Conflict of Interest}}

%\noindent {\it{The authors have declared no conflict of interest.}}

%Please insert appendices before the references. 

\bibliography{references}

\begin{thebibliography}{}

\bibitem[Aalen et~al., 2015]{aalen2015does}
Aalen, O.~O., Cook, R.~J., and R{\o}ysland, K. (2015).
\newblock Does cox analysis of a randomized survival study yield a causal treatment effect?
\newblock {\em Lifetime data analysis}, 21:579--593.

\bibitem[Alemayehu et~al., 2018]{alemayehu2018}
Alemayehu, D., Chen, Y., and Markatou, M. (2018).
\newblock A comparative study of subgroup identification methods for differential treatment effect: performance metrics and recommendations.
\newblock {\em Statistical Methods in Medical Research}, 27(12):3658--3678.

\bibitem[Amatya et~al., 2021]{amatya2021subgroup}
Amatya, A.~K., Fiero, M.~H., Bloomquist, E.~W., Sinha, A.~K., Lemery, S.~J., Singh, H., Ibrahim, A., Donoghue, M., Fashoyin-Aje, L.~A., de~Claro, R.~A., et~al. (2021).
\newblock Subgroup analyses in oncology trials: regulatory considerations and case examples.
\newblock {\em Clinical Cancer Research}, 27(21):5753--5756.

\bibitem[Arora et~al., 2021]{arora2021fda}
Arora, S., Balasubramaniam, S., Zhang, H., Berman, T., Narayan, P., Suzman, D., Bloomquist, E., Tang, S., Gong, Y., Sridhara, R., et~al. (2021).
\newblock Fda approval summary: olaparib monotherapy or in combination with bevacizumab for the maintenance treatment of patients with advanced ovarian cancer.
\newblock {\em The Oncologist}, 26(1):e164--e172.

\bibitem[Brookes et~al., 2004]{brookes2004}
Brookes, S.~T., Whitely, E., Egger, M., Smith, G.~D., Mulheran, P.~A., and Peters, T.~J. (2004).
\newblock Subgroup analyses in randomized trials: risks of subgroup-specific analyses;: power and sample size for the interaction test.
\newblock {\em Journal of clinical epidemiology}, 57(3):229--236.

\bibitem[Chen et~al., 2015]{chen2015}
Chen, G., Zhong, H., Belousov, A., and Devanarayan, V. (2015).
\newblock A prim approach to predictive‐signature development for patient stratification.
\newblock {\em Statistics in medicine}, 34(2):317--342.

\bibitem[Chernozhukov et~al., 2018]{chernozhukov2018}
Chernozhukov, V., Demirer, M., Duflo, E., and Fernandez-Val, I. (2018).
\newblock Generic machine learning inference on heterogeneous treatment effects in randomized experiments, with an application to immunization in india.
\newblock {\em National Bureau of Economic Research}.

\bibitem[Daniel et~al., 2021]{daniel2021making}
Daniel, R., Zhang, J., and Farewell, D. (2021).
\newblock Making apples from oranges: Comparing noncollapsible effect estimators and their standard errors after adjustment for different covariate sets.
\newblock {\em Biometrical Journal}, 63(3):528--557.

\bibitem[Dusseldorp et~al., 2010]{dusseldorp2010}
Dusseldorp, E., Conversano, C., and Van~Os, B. (2010).
\newblock Combining an additive and tree-based regression model simultaneously: Stima.
\newblock {\em Journal of Computational and Graphical Statistics}, 19(3):514--530.

\bibitem[{European medicines agency}, 2019]{ema2019}
{European medicines agency} (2019).
\newblock Guideline on the investigation of subgroups in confirmatory clinical trials.

\bibitem[Grouin et~al., 2005]{grouin2005}
Grouin, J.-M., Coste, M., and Lewis, J. (2005).
\newblock Subgroup analyses in randomized clinical trials: statistical and regulatory issues.
\newblock {\em Journal of biopharmaceutical statistics}, 15(5):869--882.

\bibitem[Hern{\'a}n, 2010]{hernan2010hazards}
Hern{\'a}n, M.~A. (2010).
\newblock The hazards of hazard ratios.
\newblock {\em Epidemiology (Cambridge, Mass.)}, 21(1):13.

\bibitem[Hothorn and Zeileis, 2015]{hothorn2015partykit}
Hothorn, T. and Zeileis, A. (2015).
\newblock partykit: A modular toolkit for recursive partytioning in r.
\newblock {\em The Journal of Machine Learning Research}, 16(1):3905--3909.

\bibitem[Huang et~al., 2017]{huang2017}
Huang, X., Sun, Y., Trow, P., Chatterjee, S., Chakravartty, A., Tian, L., and Devanarayan, V. (2017).
\newblock Patient subgroup identification for clinical drug development.
\newblock {\em Statistics in medicine}, 36(9):1414--1428.

\bibitem[Huber et~al., 2019]{huber2019}
Huber, C., Benda, N., and Friede, T. (2019).
\newblock A comparison of subgroup identification methods in clinical drug development: Simulation study and regulatory considerations.
\newblock {\em Pharmaceutical statistics}, 18(5):600--626.

\bibitem[Imai and Ratkovic, 2013]{imai2013}
Imai, K. and Ratkovic, M. (2013).
\newblock Estimating treatment effect heterogeneity in randomized program evaluation.
\newblock {\em Project Euclid}.

\bibitem[Imbens and Rubin, 2015]{imbens2015causal}
Imbens, G.~W. and Rubin, D.~B. (2015).
\newblock {\em Causal inference in statistics, social, and biomedical sciences}.
\newblock Cambridge University Press.

\bibitem[K{\"u}nzel et~al., 2019]{kunzel2019metalearners}
K{\"u}nzel, S.~R., Sekhon, J.~S., Bickel, P.~J., and Yu, B. (2019).
\newblock Metalearners for estimating heterogeneous treatment effects using machine learning.
\newblock {\em Proceedings of the national academy of sciences}, 116(10):4156--4165.

\bibitem[Larsen et~al., 2022]{larsen2022statistical}
Larsen, N., Stallrich, J., Sengupta, S., Deng, A., Kohavi, R., and Stevens, N. (2022).
\newblock Statistical challenges in online controlled experiments: a review of a/b testing methodology.
\newblock {\em arXiv preprint arXiv:2212.11366}.

\bibitem[LeBlanc et~al., 2005]{leblanc2005}
LeBlanc, M., Moon, J., and Crowley, J. (2005).
\newblock Adaptive risk group refinement.
\newblock {\em Biometrics}, 61(2):370--378.

\bibitem[Lipkovich et~al., 2011]{lipkovich2011}
Lipkovich, I., Dmitrienko, A., Denne, J., and Enas, G. (2011).
\newblock Subgroup identification based on differential effect search—a recursive partitioning method for establishing response to treatment in patient subpopulations.
\newblock {\em Statistics in medicine}, 30(21):2601--2621.

\bibitem[Loh, 2002]{loh2002}
Loh, W. (2002).
\newblock Regression trees with unbiased variable selection and interaction detection.
\newblock {\em Statistica sinica}, pages 361--386.

\bibitem[Loh et~al., 2019]{loh2019}
Loh, W., Cao, L., and Zhou, P. (2019).
\newblock Subgroup identification for precision medicine: A comparative review of 13 methods.
\newblock {\em Wiley Interdisciplinary Reviews: Data Mining and Knowledge Discovery}, 9(5).

\bibitem[Loh, 2011]{loh2011classification}
Loh, W.-Y. (2011).
\newblock Classification and regression trees.
\newblock {\em Wiley interdisciplinary reviews: data mining and knowledge discovery}, 1(1):14--23.

\bibitem[Patel et~al., 2016]{patel2016}
Patel, S., Hee, S., Mistry, D., Jordan, J., Brown, S., Dritsaki, M., Ellard, D., Friede, T., Lamb, S., Lord, J., and Madan, J. (2016).
\newblock Identifying back pain subgroups: developing and applying approaches using individual patient data collected within clinical trials.
\newblock {\em Programme Grants for Applied Research}, 4(10):111--134.

\bibitem[{Python Core Team}, 2020]{python}
{Python Core Team} (2020).
\newblock {\em Python Language Reference, version 3.9}.
\newblock Python Software Foundation.

\bibitem[{R Core Team}, 2021]{rsoftware}
{R Core Team} (2021).
\newblock {\em R: A Language and Environment for Statistical Computing}.
\newblock R Foundation for Statistical Computing, Vienna, Austria.

\bibitem[Ray-Coquard et~al., 2019]{ray2019olaparib}
Ray-Coquard, I., Pautier, P., Pignata, S., P{\'e}rol, D., Gonz{\'a}lez-Mart{\'\i}n, A., Berger, R., Fujiwara, K., Vergote, I., Colombo, N., M{\"a}enp{\"a}{\"a}, J., et~al. (2019).
\newblock Olaparib plus bevacizumab as first-line maintenance in ovarian cancer.
\newblock {\em New England Journal of Medicine}, 381(25):2416--2428.

\bibitem[Scornet, 2023]{scornet2023trees}
Scornet, E. (2023).
\newblock Trees, forests, and impurity-based variable importance in regression.
\newblock In {\em Annales de l'Institut Henri Poincare (B) Probabilites et statistiques}, volume 59-1, pages 21--52. Institut Henri Poincar{\'e}.

\bibitem[Sechidis et~al., 2018]{sechidis2018}
Sechidis, K., Papangelou, K., Metcalfe, P., Svensson, D., Weatherall, J., and Brown, G. (2018).
\newblock Distinguishing prognostic and predictive biomarkers: an information theoretic approach.
\newblock {\em Bioinformatics}, 34(19):3365--3376.

\bibitem[Seibold et~al., 2016]{seibold2016}
Seibold, H., Zeileis, A., and Hothorn, T. (2016).
\newblock Model-based recursive partitioning for subgroup analyses.
\newblock {\em The international journal of biostatistics.}, 12(1):45--63.

\bibitem[Su et~al., 2008]{su2008}
Su, X., Zhou, T., Yan, X., Fan, J., and Yang, S. (2008).
\newblock Interaction trees with censored survival data.
\newblock {\em The international journal of biostatistics}, 4(1).

\bibitem[Sun et~al., 2022]{sun2022}
Sun, S., Sechidis, K., Chen, Y., Lu, J., Ma, C., Mirshani, A., Ohlssen, D., Vandemeulebroecke, M., and Bornkamp, B. (2022).
\newblock Comparing algorithms for characterizing treatment effect heterogeneity in randomized trials.
\newblock {\em Biometrical Journal}.

\bibitem[Tanniou et~al., 2016]{tanniou2016}
Tanniou, J., Van Der~Tweel, I., Teerenstra, S., and Roes, K.~C. (2016).
\newblock Subgroup analyses in confirmatory clinical trials: time to be specific about their purposes.
\newblock {\em BMC medical research methodology}, 16:1--15.

\bibitem[Ternes et~al., 2017]{ternes2017}
Ternes, N., Rotolo, F., Heinze, G., and Michiels, S. (2017).
\newblock Identification of biomarker‐by‐treatment interactions in randomized clinical trials with survival outcomes and high‐dimensional spaces.
\newblock {\em Biometrical Journal}, 59(4):685--701.

\bibitem[Tian and Tibshirani, 2011]{tian2011}
Tian, L. and Tibshirani, R. (2011).
\newblock Adaptive index models for marker-based risk stratification.
\newblock {\em Biostatistics}, 12(1):68--86.

\bibitem[Wager and Athey, 2018]{wager2018}
Wager, S. and Athey, S. (2018).
\newblock Estimation and inference of heterogeneous treatment effects using random forests.
\newblock {\em Journal of the American Statistical Association}, 113(523):1228--1242.

\bibitem[Xie et~al., 2018]{xie2018false}
Xie, Y., Chen, N., and Shi, X. (2018).
\newblock False discovery rate controlled heterogeneous treatment effect detection for online controlled experiments.
\newblock In {\em Proceedings of the 24th ACM SIGKDD international conference on knowledge discovery \& data mining}, pages 876--885.

\bibitem[Xu et~al., 2023]{xu2023treatment}
Xu, Y., Ignatiadis, N., Sverdrup, E., Fleming, S., Wager, S., and Shah, N. (2023).
\newblock Treatment heterogeneity with survival outcomes.
\newblock In {\em Handbook of Matching and Weighting Adjustments for Causal Inference}, pages 445--482. Chapman and Hall/CRC.

\bibitem[Xu et~al., 2015]{xu2015}
Xu, Y., Yu, M., Zhao, Y., Li, Q., Wang, S., and Shao, J. (2015).
\newblock Regularized outcome weighted subgroup identification for differential treatment effects.
\newblock {\em Biometrics}, 71(3):645--653.

\bibitem[Zeileis et~al., 2008]{zeileis2008model}
Zeileis, A., Hothorn, T., and Hornik, K. (2008).
\newblock Model-based recursive partitioning.
\newblock {\em Journal of Computational and Graphical Statistics}, 17(2):492--514.

\bibitem[Zhao et~al., 2012]{zhao2012}
Zhao, Y., Zeng, D., Rush, A., and Kosorok, M. (2012).
\newblock Estimating individualized treatment rules using outcome weighted learning.
\newblock {\em Journal of the American Statistical Association}, 107(499):1106--1118.

\end{thebibliography}

\newpage

\appendix
\renewcommand{\thesubsection}{A.\arabic{subsection}}

\section*{Appendix}

\subsection{Research questions and methods}
\label{appendix:methods}

We selected 9 methods for our benchmark, based on existing implementation, types of outcomes handled and applicable research questions. 
A summary of these characteristics for identified methods in available in \cref{table:methods1} below.

\begin{table}[htb]
\begin{center}
\caption{Summary of existing subgroup analysis methods, including availability of answers to research questions by default and available implementation.}
\label{table:methods1}
\begin{tabular}{llllll}
\hline
 \multirow{2}{*}{Methods}&
  \multicolumn{3}{l}{Research questions} &
  \multicolumn{2}{l}{Existing implementation} \\
  & 1 & 2 & 3 & Language & Time-to-event \\
%Description 1 & Description 2 & Description 3\\
\hline
 Univariate Interaction & \cmark & \cmark & \xmark & Python\footref{note:blocks} & \cmark \\
 Univariate t-test &  \cmark  & \cmark & \xmark & Python\footref{note:blocks} & \cmark \\
 Multivariate Cox & \cmark & \cmark & \xmark & Python\footref{note:blocks} & \cmark \\
 Multivariate Tree & \cmark & \xmark & \cmark & Python\footref{note:blocks} & \cmark \\
 MOB & \cmark & \cmark & \cmark & R & \cmark \\
 IT & \cmark & \cmark &  \cmark & \xmark & \xmark \\
 SIDES & \cmark & \cmark & \cmark & R & \cmark \\
 SeqBT & \cmark & \cmark & \xmark & \xmark & \xmark \\
 ARDP & \xmark & \cmark & \xmark & \xmark & \xmark \\
\hline
\end{tabular}
\end{center}
\end{table}

\noindent Methods differ from each other by several aspects; in particular they do not all provide similar ways of answering to the research questions of interest as discussed in Sections \ref{method_rq1}, \ref{method_rq2} and \ref{method_rq3}. Here we provide more information on testing the existence of treatment effect heterogeneity for predictive methods (Section \ref{method_rq1}) and we also discuss whether there is an existing implementation for the selected methods.

\medskip

\noindent Predictive methods are the methods for which the existence of heterogeneity is assessed once patient subgroups are formed: the multivariate methods, SIDES, SeqBT, ARDP, as well as the univariate t-test method which makes use of the same principle. For these methods, the research question \ref{question:1} is answered by performing a statistical test between the two predicted subgroups. The statistical test consists of computing a difference-in-differences test:
\begin{equation}\label{eq:test_pval}
\frac{ (\hat{\mu}_{01} - \hat{\mu}_{00}) - (\hat{\mu}_{11} - \hat{\mu}_{10})}{\hat{\sigma}_{01}^2 + \hat{\sigma}_{00}^2 + \hat{\sigma}_{11}^2 + \hat{\sigma}_{10}^2},
\end{equation}
where $\hat{\mu}_{ij}$ (resp. $\hat{\sigma}_{ij}^2$) is an estimator of the median (resp. variance) of the times to event of the patients in subgroup $G=i$ with $W=j$.

\medskip

\noindent A few subgroup analysis methods have building blocks implemented in Python\footnote{\label{note:blocks} Available building blocks generally only include the model fitting step; we extended their implementation to adapt it to our full benchmarking pipeline and provide answers to each research question, using necessary adaptation}. In particular, the univariate methods are implemented for each covariate by fitting a Cox model using the \texttt{lifelines} package; two different multivariate methods are trained, a multivariate Cox model using \texttt{lifelines} and a tree-based multivariate model using \texttt{scikit-survival}. 
Some other methods are implemented but only in R, such as MOB, implemented in the \texttt{partykit} package \citep{hothorn2015partykit}; SIDES in the \texttt{SIDES} package; and SeqBT implemented in the \texttt{SubgrpID} package.
Finally the other methods (ARDP, ITree) are not publicly available in any programming language.

\subsection{Data generating process}
\label{appendix:dgp}

Recall that we take the following hazard function to generate data:
\begin{equation*}\label{eq:our_synthetic_model}
h(t \, | \, \mathbf{X}, W) = h_0(t) \exp\left( \beta_0 W + (\beta_1 - \beta_0) G(X) W + \gamma^T X \right),
\end{equation*}
where:
\begin{itemize}
    \item $h_0(t)$ is a baseline hazard function, which is a function depending only on $t$ that takes non-negative values and that satisfies $\int_0^{+\infty} h_0(u) \mathrm{d}u = +\infty$,
    \item $G: \mathbb{R}^p \rightarrow \{0, 1\}$ is a subgroup function that outputs $0$ or $1$ from covariates $\bf X$,
    \item $\beta_0:=\log\left( \frac{h(t \, | \, \mathbf{X}, W=1, G=0)}{h(t \, | \, \mathbf{X}, W=0, G=0)} \right)$ is the log-hazard ratio in subgroup $G=0$,
    \item $\beta_1:=\log\left( \frac{h(t \, | \, \mathbf{X}, W=1, G=1)}{h(t \, | \, \mathbf{X}, W=0, G=1)} \right)$ is the log-hazard ratio in subgroup $G=1$,
    \item $\gamma \in \mathbb{R}^p$ is a prognostic vector.
\end{itemize}

\medskip

\noindent {\bf Monte-Carlo estimation.} Fix $\beta \in \mathbb{R}$. Let $N \geq 1$ be a large integer. Our Monte-Carlo estimation of the quantities $\ARR_0(t, \beta_0=\beta)$ and $\ARR_1(t, \beta_1=\beta)$ goes as follows. For all $1 \leq i \leq N$:
\begin{enumerate}
    \item Generate $\mathbf{X}_i$ from its law,
    \item Compute subgroup identity $G(\mathbf{X}_i)$,
    \item Compute the ground truth individual ARR: $\ARR_i(t) = S(t \, | \, \mathbf{X}, W=1) - S(t \, | \, \mathbf{X}, W=0)$.
\end{enumerate}
Finally, denoting by $\mathcal{I}_0 = \{1 \leq i \leq N, \, G(\mathbf{X}_i)=0 \}$ and $\mathcal{I}_1 = \{1 \leq i \leq N, \, G(\mathbf{X}_i)=1 \}$, compute the Monte-Carlo estimates:
\[ \widehat{\ARR}_0(t, \beta_0=\beta) = \frac{1}{|\mathcal{I}_0|} \sum\limits_{i \in \mathcal{I}_0} \ARR_i(t) \, \, ; \, \, \widehat{\ARR}_1(t, \beta_1=\beta) = \frac{1}{|\mathcal{I}_1|} \sum\limits_{i \in \mathcal{I}_1} \ARR_i(t).  \]
The idea is to repeat the procedure for a refined grid of $\beta$'s values. We took $N=10^6$ in our simulations. By the law of large numbers, the estimation error is of order $1/\sqrt{N} = 0.001$. The results of this Monte-Carlo procedure can then be used for the benchmark: for a given amount of heterogeneity, measured as a difference in $\ARR$ between subgroup $G=0$ and subgroup $G=1$, we are able to choose the exact corresponding values of the parameters $\beta_0$ and $\beta_1$, and generate the data accordingly. 

\smallskip

\noindent {\bf Remark.} {\it The functions $\beta_0 \mapsto \ARR(t=1 \, | \, G=0)$ and $\beta_1 \mapsto \ARR(t=1 \, | \, G=1)$ are one-to-one, which enables to unambiguously choose $\beta_0$ and $\beta_1$. Indeed, using the fact that $S(t) = \exp( - \int h(u) \mathrm{d}u )$, the quantity $\ARR(t=1 \, | \, G=0)$ can be written as:
\[ \frac{1}{\mathbb{P}(G=0)} \left[ \exp\left( - \exp( \beta_0 + \gamma^T X ) \int_{t=0}^{t=1}  h_0(u) \mathrm{d}u \right) - \exp\left( - \exp( \gamma^T X ) \int_{t=0}^{t=1} h_0(u) \mathrm{d}u \right) \right], \]
which is a non-increasing function of $\beta_0$. The same argument holds for $\ARR(t=1 \, | \, G=1)$.}

\smallskip

\noindent {\bf Generation step.} Recall that we compute the $\ARR$ at time $t=1$ and focus on overall non-significant trials, meaning that $\ARR(1) = 0$. In term of the ARRs in the two subgroups, this translates into 
\begin{equation}\label{eq:ARR_relation}
\ARR(1; \,  G=0)\mathbb{P}(G=0) + \ARR(1; \, G=1)\mathbb{P}(G=1) = 0.
\end{equation} 
Hence, we vary ARR in subgroup $G=1$ and obtain the ARR in subgroup $G=0$ from \eqref{eq:ARR_relation}. By default, we consider $G=1$ as the subgroup of good responders and make the ARR vary from $0$, which corresponds to the null hypothesis where there is no heterogeneity ($\beta_0=\beta_1=0$), to a maximum value that is constrained by \eqref{eq:ARR_relation} and the range of values taken by the functions $\beta_0 \mapsto \ARR(t \, | \, G=0)$ and $\beta_1 \mapsto \ARR(1 \, | \, G=1)$. In our experiments, we always take $10$ evenly distributed $\ARR$s between $\ARR(1; G=1) = 0$ and the maximal achievable value of $\ARR(1; G=1)$ given the constraint \eqref{eq:ARR_relation}.

\medskip

\noindent {\bf Our scenarios choices.} We consider three values of dimension $p=20$, $100$ and $1000$, and always take $\mathbf{X} \sim \mathcal{N}(0_p, I_p)$ following an isotropic Gaussian distribution. We consider a Weibull baseline hazard function $h_0(t) = t$. The prognostic vector $\gamma$ is chosen so as to obtain a sparse prognostic structure. In dimension $20$,
\begin{equation}\label{eq: gamma20}
\gamma_i = \left\{
\begin{array}{rr}
    1  & \text{if $1 \leq i \leq 5$,}  \\
    -1 & \text{if $6 \leq i \leq 10$,} \\
    0 & \text{if $11 \leq i \leq 20$,} \\
\end{array}
\right.
\end{equation}
in dimension $100$:
\begin{equation}\label{eq: gamma100}
\gamma_i = \left\{
\begin{array}{rl}
    1  & \text{if $1 \leq i \leq 5$,}  \\
    -1 & \text{if $6 \leq i \leq 10$,} \\
    0.1 & \text{if $11 \leq i \leq 15$ or $26 \leq i \leq 30$,} \\
    -0.1 & \text{if $16 \leq i \leq 25$,} \\
    0.01 & \text{if $31 \leq i \leq 35$ or $46 \leq i \leq 50$ or $56 \leq i \leq 60$,} \\
    -0.01 & \text{if $36 \leq i \leq 45$ or $51 \leq i \leq 55$,} \\
    0 & \text{if $61 \leq i \leq 100$,} \\
\end{array}
\right.
\end{equation}
and in dimension $1000$:
\begin{equation}\label{eq: gamma1000}
\gamma_i = \left\{
\begin{array}{rr}
    1  & \text{if $1 \leq i \leq 10$,}  \\
    -1 & \text{if $11 \leq i \leq 20$,} \\
    0.1 & \text{if $21 \leq i \leq 70$,} \\
    -0.1 & \text{if $71 \leq i \leq 120$,} \\
    0.01 & \text{if $121 \leq i \leq 320$,} \\
    -0.01 & \text{if $321 \leq i \leq 520$,} \\
    0 & \text{if $521 \leq i \leq 1000$.} \\
\end{array}
\right.
\end{equation}

In order to mimic biologically relevant subgroup functions $G$, we make the hypothesis that few biomarkers are predictive, and that they interact with the treatment when exceeding a predefined value. All of our subgroup functions are of the form 
\begin{equation}
G(x) = \mathbf{1}_{x_i \geq -1, x_j \geq -1, x_k \geq -1, x_l \geq -1 }, 
\end{equation}
for some indexes $i, j, k, l$, whose choices allow to vary the number of biomarkers that are both predictive and prognostic. In dimension $p=20$, we explore the five possible corresponding scenarios, by taking $i, j, k, l$ as follows: $(i,j,k,l)=(17,18,19,20)$, the four predictive variable are not prognostic ; $(i,j,k,l)=(9,13,14,15)$, one of the predictive variables is prognostic ; $(i,j,k,l)=(9,10,11,12)$, two of the predictive variables are prognostic ; $(i,j,k,l)=(6,7,8,20)$, three of the predictive variables are prognostic ; and $(i,j,k,l)=(1,2,7,8)$, all predictive variables are prognostic. In dimension $p=100$ and $p=1000$, we only consider the setting where the predictive biomarkers are not prognostic. 

Since $\mathbf{X}$ is an isotropic gaussian variable $\mathbb{P}(G=1) \approx 1/2$ and the subgroups of good and bad responders are balanced. Although we restrict our considerations to the setting \eqref{eq: subgroup_defn}, note that our code pipeline allows us to use any choice of function $G$.

Finally, we explore several sampling sizes $N=100, 500, 1000$ as well as train/test splitting: $0.25/0.75$, $0.5/0.5$ and $0.75/0.25$.

\medskip

\noindent {\bf Censoring.} Let $N \geq 1$. Let $(T_i, W_i, \mathbf{X}_i)_{1 \leq i \leq N}$ be generated as before. We would like to add censoring to the data in order to mimic real-life clinical trials. We generated another independent random variable $C \sim \mathcal{B}(a,b)$, a beta law, re-scaled on $[0,20]$. Then, we define $U=\min(T,C)$ and say that censoring occurs if and only if $U<T$. We further define $E=\mathbf{1}_{T\leq U}$, the indicator that an event is observed. We then only get to observe the censored version $(U_i, E_i, W_i, \mathbf{X}_i)_{1 \leq i \leq N}$ of the dataset. In order to investigate three increasing censoring scales, we consider three pairs of values for the parameters $a$ and $b$ of the beta law: $(0.4,0.4)$, $(0.3, 1.0)$ and $(0.2,2.0)$. For the subgroup definitions with $4$ predictive biomarkers that are not prognostic, and for a value $\ARR(t=1; \, G=1)=0.2$, the corresponding event rates ($E=1$) are reported in Table \ref{table:censorship}.

\begin{table}[htb]
\begin{center}
\caption{Event rates for the three censoring scenarios. We always take the subgroup definition with $4$ biomarkers that are not prognostic, and compute the rate at $\ARR(1; \, G=1)=0.2$.}
\label{table:censorship}
\begin{tabular}{c|cc|ccc}
Scenario & a & b & p=20 & p=100 & p=1000 \\
\hline
(1) & 0.4 & 0.4 & 77\% & 81\% & 70\% \\
(2) & 0.3 & 1.0 & 62\% & 56\% & 55\% \\
(3) & 0.2 & 2.0 & 36\% & 40\% & 39\% \\
\hline
\end{tabular}
\end{center}
\end{table}

\subsection{Reproducibility}

We provide below all necessary information to reproduce our analyses. Our data generation process, partially based on pseudorandom number generation, is seed-controlled to ensure reproducibility. 
%However, our experiments are run in parallel processes to computation efficiency, as such, there are some floating point arithmetic issues and non-determinism. 
This seed control, as well as the 100 repetitions per experiment to average the differences out, enable reproducibility of the results.
Reproducing one experiment for a set of parameters requires 3 main steps: {\it(i)} the definition of parameters and experiment settings; {\it(ii)} the generation of the data using the DGP; {\it(iii)} running the experiments.
Every reproducibility experiment can be run from the \texttt{subgroup\_analysis} repository, which contains complementary and detailed information in the README file to perform the three steps described below.

\medskip

\noindent{\textbf{Definition of parameters and experiment settings.}} This step consists of establishing the parameters of the experiment to be run, more specifically the parameters of the data generation process, which are {\it(i)} the definition of the subgroup of interest; {\it(ii)} the number of covariates to be considered; {\it(iii)} the characterization of the covariates (means, covariance structure and prognostic values). Subgroups were defined to be balanced, to use 4 predictive variables and varying number of prognostic variables (from 0 to 4 prognostic variables). For the fully synthetic data, 3 dimensions are studied, with 20, 100 and 1000 covariates. More specifically, the prognostic vector for the dimension $p=20$ is available in \cref{sec:dgp}. The prognostic vector in dimension 100 is

\medskip

\noindent{\textbf{Generation of the data using the DGP.}} Once parameters are set, the DGP is used to produce the range of $\ARR(1; \,  G=0)$ and $\ARR(1; \,  G=1)$ across $\beta = [-10;10]$ -- refer to \cref{sec:dgp} for more details. The DGP is used based on the parameters defined in the previous step, a subgroup definition, and the $\beta$ range.

\medskip

\noindent{\textbf{Run of experiments.}} When the data has been generated, experiments can be run. Each experiment consist of 10 points of ARR, between 0 and $\max(\max(\ARR(1; \,  G=1)), \min(\ARR(1; \,  G=0)))$. At each point of ARR, the experiment is run 100 times for metrics stability. For a summary of all experiments launched, see \cref{table:params} below. Each experiment is defined by the data it uses; the range of ARR points considered; the sample, training and test sizes; and the censoring scenario. Refer to \cref{appendix:dgp} for details about the three censoring scenarios.

\begin{table}[htb]
\begin{center}
\caption{Summary of parameters used for experiments}
\label{table:params}
\begin{tabular}{c|c}
Categories & Parameters \\
\hline
Prognostic predictive variables & 0, 1, 2, 3, 4 \\
Sample size & 100, 500, 1000 \\
Training set proportion & 0.25, 0.5, 0.75 \\
Censoring & False, True (scenarios 1, 2, 3) \\
\hline
\end{tabular}
\end{center}
\end{table}

%\noindent{\textbf{Analysis of experiments.}} Experiments results can be analyzed in the research question framework using Jupyter notebooks. In particular, the type I error and power are computed, as well as precision average score. All computed metrics, as well as other metrics available from the experiments (variables ranking, accuracy and other classification metrics) can be analyzed with plots. The figures presented in this paper can be reproduced using the notebooks available in \texttt{subgroup\_analysis}.

\subsection{Benchmarking experiments}
\label{app:benchmark}

\subsubsection{Complexity of methods and increasing dimensions}

\begin{figure}[htb]
\begin{center}
\includegraphics[scale=0.45]{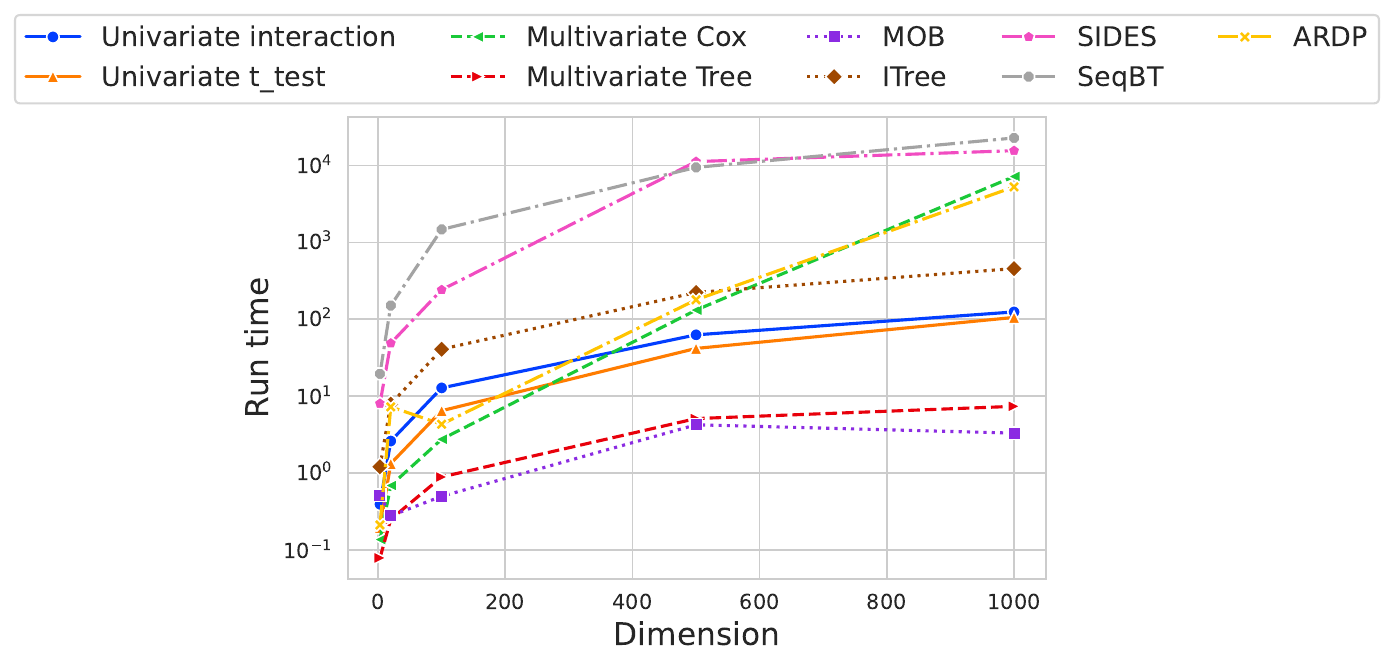}
\caption{Run time (in seconds) of each method fitting step for dimensions p=3, p=20, p=100, p=500 and p=1000, in log-scale.}
\label{fig:complexity}
\end{center}
\end{figure}

Subgroup analysis methods differ in terms of complexity, and the time required to run each evolves with the dimension of the covariate space. Some methods are better suited for large dimensions, while others are strongly limited by their complexity. To benchmark these methods in different dimension scenarios, establishing the experiments' run times is of importance. As such, we empirically assessed the behavior of subgroup analysis methods when the dimension of the covariate space increases by measuring the duration of the fitting step of each method -- see \cref{fig:complexity}. The MOB and multivariate tree methods scale very well to large dimension; SIDES and SeqBT fitting is very inefficient in high dimensions; other methods run times are intermediate. Given these results, SIDES and SeqBT were excluded from high dimension benchmarking experiments (p=100 and p=1000) and are considered to be better suited for small covariate spaces.

\subsubsection{Research questions}
\label{app:rqs_table}
The remaining of this supplementary material is made of a detailed report of our results. For research question \ref{question:1}, we report tables of statistical power. For research question \ref{question:2}, we report tables of the averaged precision scores. which is the area under the precision-recall curve. Finally, for research question \ref{question:3}, we report tables of the accuracy. Each table contains all methods, four ARR points and each censoring scenarios. The experiments used 100 repeats of the DGP under the null hypothesis, with 500 samples and a 0.5/0.5 split between training and testing sets. Notice that we have in fact more results that can be investigated with the attached file. More precisely, the interested reader may find the following csv files results:
\begin{enumerate}
    \item {\bf Type I error.} For $p=20$ (5 subgroup definitions), $p=100$ and $p=1000$, we computed refined type I error estimates for the three censoring scenarios, using 1000 repetitions of the DGP at ARR=0.
    \item {\bf Sampling size 500, train/test split 0.5/0.5.} In this context, we experiment on each subgroup definitions in dimensions $p=20, 100, 1000$, without censoring and for the three censoring scenarios. 
    \item {\bf Sampling size 1000, train/test split 0.5/0.5.} In this context, we experiment on each subgroup definitions in dimensions $p=20, 100, 1000$, without censoring and for the three censoring scenarios. 
    \item {\bf Miscellaneous.} In dimension $p=20$, for subgroup definitions with 0 and 1 variables we also investigated, without censoring and for the three censoring scenarios, sampling sizes $100, 500, 1000$ with three train/test split choices: 0.25/0.75, 0.5/0.5 and 0.75/0.25.
\end{enumerate}

\begin{table}[!htb]
\begin{center}
\caption{Power analysis of each method for p=20 for the subgroup definition with 4 predictive variables that are not prognostic.}
\label{sup:prog0_power}
\begin{tabular}{llcccc}
\hline
arr & method & 0 & 1 & 2 & 3   \\
\hline
\multirow[t]{10}{*}{0.000000} & ARDP & 0.08 $\pm$ 0.05 & 0.06 $\pm$ 0.05 & 0.09 $\pm$ 0.06 & 0.03 $\pm$ 0.03 \\
 & ITree & 0.07 $\pm$ 0.05 & 0.02 $\pm$ 0.03 & 0.02 $\pm$ 0.03 & 0.00 $\pm$ 0.00 \\
 & MOB & 0.05 $\pm$ 0.04 & 0.07 $\pm$ 0.05 & 0.03 $\pm$ 0.03 & 0.03 $\pm$ 0.03 \\
 & Multivariate cox & 0.06 $\pm$ 0.05 & 0.05 $\pm$ 0.04 & 0.02 $\pm$ 0.03 & 0.07 $\pm$ 0.05 \\
 & Multivariate tree & 0.05 $\pm$ 0.04 & 0.05 $\pm$ 0.04 & 0.05 $\pm$ 0.04 & 0.05 $\pm$ 0.04 \\
 & Oracle & 0.01 $\pm$ 0.02 & 0.02 $\pm$ 0.03 & 0.03 $\pm$ 0.03 & 0.02 $\pm$ 0.03 \\
 & SIDES & 0.14 $\pm$ 0.07 & 0.10 $\pm$ 0.06 & 0.03 $\pm$ 0.03 & 0.15 $\pm$ 0.07 \\
 & SeqBT & 0.05 $\pm$ 0.04 & 0.05 $\pm$ 0.04 & 0.04 $\pm$ 0.04 & 0.01 $\pm$ 0.02 \\
 & Univariate interaction & 0.02 $\pm$ 0.03 & 0.05 $\pm$ 0.04 & 0.02 $\pm$ 0.03 & 0.01 $\pm$ 0.02 \\
 & Univariate t-test & 0.03 $\pm$ 0.03 & 0.03 $\pm$ 0.03 & 0.05 $\pm$ 0.04 & 0.01 $\pm$ 0.02 \\
\cline{1-6}
\multirow[t]{10}{*}{0.144137} & ARDP & 0.08 $\pm$ 0.05 & 0.04 $\pm$ 0.04 & 0.00 $\pm$ 0.00 & 0.04 $\pm$ 0.04 \\
 & ITree & 0.16 $\pm$ 0.07 & 0.07 $\pm$ 0.05 & 0.07 $\pm$ 0.05 & 0.01 $\pm$ 0.02 \\
 & MOB & 0.12 $\pm$ 0.06 & 0.13 $\pm$ 0.07 & 0.12 $\pm$ 0.06 & 0.07 $\pm$ 0.05 \\
 & Multivariate cox & 0.04 $\pm$ 0.04 & 0.08 $\pm$ 0.05 & 0.05 $\pm$ 0.04 & 0.04 $\pm$ 0.04 \\
 & Multivariate tree & 0.07 $\pm$ 0.05 & 0.07 $\pm$ 0.05 & 0.07 $\pm$ 0.05 & 0.07 $\pm$ 0.05 \\
 & Oracle & 0.88 $\pm$ 0.06 & 0.90 $\pm$ 0.06 & 0.74 $\pm$ 0.09 & 0.60 $\pm$ 0.10 \\
 & SIDES & 0.15 $\pm$ 0.07 & 0.14 $\pm$ 0.07 & 0.10 $\pm$ 0.06 & 0.15 $\pm$ 0.07 \\
 & SeqBT & 0.01 $\pm$ 0.02 & 0.07 $\pm$ 0.05 & 0.07 $\pm$ 0.05 & 0.04 $\pm$ 0.04 \\
 & Univariate interaction & 0.08 $\pm$ 0.05 & 0.08 $\pm$ 0.05 & 0.07 $\pm$ 0.05 & 0.02 $\pm$ 0.03 \\
 & Univariate t-test & 0.10 $\pm$ 0.06 & 0.05 $\pm$ 0.04 & 0.03 $\pm$ 0.03 & 0.00 $\pm$ 0.00 \\
\cline{1-6}
\multirow[t]{10}{*}{0.288274} & ARDP & 0.07 $\pm$ 0.05 & 0.03 $\pm$ 0.03 & 0.05 $\pm$ 0.04 & 0.04 $\pm$ 0.04 \\
 & ITree & 0.79 $\pm$ 0.08 & 0.65 $\pm$ 0.09 & 0.49 $\pm$ 0.10 & 0.18 $\pm$ 0.08 \\
 & MOB & 0.60 $\pm$ 0.10 & 0.52 $\pm$ 0.10 & 0.46 $\pm$ 0.10 & 0.33 $\pm$ 0.09 \\
 & Multivariate cox & 0.29 $\pm$ 0.09 & 0.20 $\pm$ 0.08 & 0.16 $\pm$ 0.07 & 0.07 $\pm$ 0.05 \\
 & Multivariate tree & 0.04 $\pm$ 0.04 & 0.05 $\pm$ 0.04 & 0.04 $\pm$ 0.04 & 0.04 $\pm$ 0.04 \\
 & Oracle & 1.00 $\pm$ 0.00 & 1.00 $\pm$ 0.00 & 1.00 $\pm$ 0.00 & 0.98 $\pm$ 0.03 \\
 & SIDES & 0.20 $\pm$ 0.08 & 0.14 $\pm$ 0.07 & 0.12 $\pm$ 0.06 & 0.16 $\pm$ 0.07 \\
 & SeqBT & 0.23 $\pm$ 0.08 & 0.26 $\pm$ 0.09 & 0.14 $\pm$ 0.07 & 0.07 $\pm$ 0.05 \\
 & Univariate interaction & 0.56 $\pm$ 0.10 & 0.53 $\pm$ 0.10 & 0.32 $\pm$ 0.09 & 0.12 $\pm$ 0.06 \\
 & Univariate t-test & 0.26 $\pm$ 0.09 & 0.18 $\pm$ 0.08 & 0.16 $\pm$ 0.07 & 0.09 $\pm$ 0.06 \\
\cline{1-6}
\multirow[t]{10}{*}{0.432411} & ARDP & 0.18 $\pm$ 0.08 & 0.19 $\pm$ 0.08 & 0.17 $\pm$ 0.07 & 0.09 $\pm$ 0.06 \\
 & ITree & 0.99 $\pm$ 0.02 & 1.00 $\pm$ 0.00 & 1.00 $\pm$ 0.00 & 0.99 $\pm$ 0.02 \\
 & MOB & 0.98 $\pm$ 0.03 & 1.00 $\pm$ 0.00 & 0.97 $\pm$ 0.03 & 0.98 $\pm$ 0.03 \\
 & Multivariate cox & 0.42 $\pm$ 0.10 & 0.60 $\pm$ 0.10 & 0.51 $\pm$ 0.10 & 0.30 $\pm$ 0.09 \\
 & Multivariate tree & 0.64 $\pm$ 0.09 & 0.57 $\pm$ 0.10 & 0.47 $\pm$ 0.10 & 0.27 $\pm$ 0.09 \\
 & Oracle & 1.00 $\pm$ 0.00 & 1.00 $\pm$ 0.00 & 1.00 $\pm$ 0.00 & 1.00 $\pm$ 0.00 \\
 & SIDES & 0.36 $\pm$ 0.09 & 0.24 $\pm$ 0.08 & 0.23 $\pm$ 0.08 & 0.18 $\pm$ 0.08 \\
 & SeqBT & 0.86 $\pm$ 0.07 & 0.83 $\pm$ 0.07 & 0.81 $\pm$ 0.08 & 0.62 $\pm$ 0.10 \\
 & Univariate interaction & 0.99 $\pm$ 0.02 & 1.00 $\pm$ 0.00 & 1.00 $\pm$ 0.00 & 0.93 $\pm$ 0.05 \\
 & Univariate t-test & 0.73 $\pm$ 0.09 & 0.69 $\pm$ 0.09 & 0.54 $\pm$ 0.10 & 0.40 $\pm$ 0.10 \\
\cline{1-6}
\hline
\end{tabular}
\end{center}
\end{table}

\begin{table}[!htb]
\begin{center}
\caption{Power analysis of each method for p=100 for the subgroup definition with 4 predictive variables that are not prognostic.}
\label{sup:prog100_power}
\begin{tabular}{llcccc}
\hline
arr & method &  0 & 1 & 2 & 3  \\
\hline
\multirow[t]{8}{*}{0.000000} & ARDP & 0.06 $\pm$ 0.05 & 0.05 $\pm$ 0.04 & 0.03 $\pm$ 0.03 & 0.06 $\pm$ 0.05 \\
 & ITree & 0.00 $\pm$ 0.00 & 0.00 $\pm$ 0.00 & 0.00 $\pm$ 0.00 & 0.00 $\pm$ 0.00 \\
 & MOB & 0.00 $\pm$ 0.00 & 0.03 $\pm$ 0.03 & 0.02 $\pm$ 0.03 & 0.03 $\pm$ 0.03 \\
 & Multivariate cox & 0.04 $\pm$ 0.04 & 0.02 $\pm$ 0.03 & 0.00 $\pm$ 0.00 & 0.04 $\pm$ 0.04 \\
 & Multivariate tree & 0.05 $\pm$ 0.04 & 0.05 $\pm$ 0.04 & 0.05 $\pm$ 0.04 & 0.05 $\pm$ 0.04 \\
 & Oracle & 0.04 $\pm$ 0.04 & 0.00 $\pm$ 0.00 & 0.08 $\pm$ 0.05 & 0.04 $\pm$ 0.04 \\
 & Univariate interaction & 0.00 $\pm$ 0.00 & 0.00 $\pm$ 0.00 & 0.01 $\pm$ 0.02 & 0.00 $\pm$ 0.00 \\
 & Univariate t-test & 0.04 $\pm$ 0.04 & 0.04 $\pm$ 0.04 & 0.07 $\pm$ 0.05 & 0.07 $\pm$ 0.05 \\
\cline{1-6}
\multirow[t]{8}{*}{0.144367} & ARDP & 0.05 $\pm$ 0.04 & 0.08 $\pm$ 0.05 & 0.01 $\pm$ 0.02 & 0.10 $\pm$ 0.06 \\
 & ITree & 0.03 $\pm$ 0.03 & 0.02 $\pm$ 0.03 & 0.00 $\pm$ 0.00 & 0.01 $\pm$ 0.02 \\
 & MOB & 0.05 $\pm$ 0.04 & 0.04 $\pm$ 0.04 & 0.08 $\pm$ 0.05 & 0.01 $\pm$ 0.02 \\
 & Multivariate cox & 0.05 $\pm$ 0.04 & 0.08 $\pm$ 0.05 & 0.04 $\pm$ 0.04 & 0.04 $\pm$ 0.04 \\
 & Multivariate tree & 0.07 $\pm$ 0.05 & 0.07 $\pm$ 0.05 & 0.07 $\pm$ 0.05 & 0.07 $\pm$ 0.05 \\
 & Oracle & 0.88 $\pm$ 0.06 & 0.83 $\pm$ 0.07 & 0.75 $\pm$ 0.09 & 0.55 $\pm$ 0.10 \\
 & Univariate interaction & 0.02 $\pm$ 0.03 & 0.01 $\pm$ 0.02 & 0.03 $\pm$ 0.03 & 0.00 $\pm$ 0.00 \\
 & Univariate t-test & 0.03 $\pm$ 0.03 & 0.04 $\pm$ 0.04 & 0.09 $\pm$ 0.06 & 0.03 $\pm$ 0.03 \\
\cline{1-6}
\multirow[t]{8}{*}{0.288733} & ARDP & 0.07 $\pm$ 0.05 & 0.10 $\pm$ 0.06 & 0.03 $\pm$ 0.03 & 0.02 $\pm$ 0.03 \\
 & ITree & 0.50 $\pm$ 0.10 & 0.51 $\pm$ 0.10 & 0.22 $\pm$ 0.08 & 0.04 $\pm$ 0.04 \\
 & MOB & 0.29 $\pm$ 0.09 & 0.30 $\pm$ 0.09 & 0.18 $\pm$ 0.08 & 0.04 $\pm$ 0.04 \\
 & Multivariate cox & 0.17 $\pm$ 0.07 & 0.07 $\pm$ 0.05 & 0.04 $\pm$ 0.04 & 0.05 $\pm$ 0.04 \\
 & Multivariate tree & 0.04 $\pm$ 0.04 & 0.04 $\pm$ 0.04 & 0.04 $\pm$ 0.04 & 0.04 $\pm$ 0.04 \\
 & Oracle & 1.00 $\pm$ 0.00 & 1.00 $\pm$ 0.00 & 1.00 $\pm$ 0.00 & 0.99 $\pm$ 0.02 \\
 & Univariate interaction & 0.25 $\pm$ 0.09 & 0.29 $\pm$ 0.09 & 0.10 $\pm$ 0.06 & 0.00 $\pm$ 0.00 \\
 & Univariate t-test & 0.09 $\pm$ 0.06 & 0.07 $\pm$ 0.05 & 0.07 $\pm$ 0.05 & 0.06 $\pm$ 0.05 \\
\cline{1-6}
\multirow[t]{8}{*}{0.433100} & ARDP & 0.09 $\pm$ 0.06 & 0.05 $\pm$ 0.04 & 0.07 $\pm$ 0.05 & 0.06 $\pm$ 0.05 \\
 & ITree & 0.98 $\pm$ 0.03 & 1.00 $\pm$ 0.00 & 0.99 $\pm$ 0.02 & 0.94 $\pm$ 0.05 \\
 & MOB & 0.95 $\pm$ 0.04 & 0.94 $\pm$ 0.05 & 0.91 $\pm$ 0.06 & 0.73 $\pm$ 0.09 \\
 & Multivariate cox & 0.11 $\pm$ 0.06 & 0.10 $\pm$ 0.06 & 0.09 $\pm$ 0.06 & 0.09 $\pm$ 0.06 \\
 & Multivariate tree & 0.54 $\pm$ 0.10 & 0.49 $\pm$ 0.10 & 0.39 $\pm$ 0.10 & 0.24 $\pm$ 0.08 \\
 & Oracle & 1.00 $\pm$ 0.00 & 1.00 $\pm$ 0.00 & 1.00 $\pm$ 0.00 & 1.00 $\pm$ 0.00 \\
 & Univariate interaction & 0.97 $\pm$ 0.03 & 1.00 $\pm$ 0.00 & 0.99 $\pm$ 0.02 & 0.75 $\pm$ 0.09 \\
 & Univariate t-test & 0.48 $\pm$ 0.10 & 0.41 $\pm$ 0.10 & 0.36 $\pm$ 0.09 & 0.16 $\pm$ 0.07 \\
\cline{1-6}
\hline
\end{tabular}
\end{center}
\end{table}

\begin{table}[!htb]
\begin{center}
\caption{Power analysis of each methods for p=1000 for the subgroup definition with 4 predictive variables that are not prognostic.}
\label{sup:prog1000_power}
\begin{tabular}{llcccc}
\hline
arr & method & 0 & 1 & 2 & 3 \\
\hline
\multirow[t]{8}{*}{0.000000} & ARDP & 0.08 $\pm$ 0.05 & 0.05 $\pm$ 0.04 & 0.06 $\pm$ 0.05 & 0.04 $\pm$ 0.04 \\
 & ITree & 0.01 $\pm$ 0.02 & 0.01 $\pm$ 0.02 & 0.01 $\pm$ 0.02 & 0.00 $\pm$ 0.00 \\
 & MOB & 0.08 $\pm$ 0.05 & 0.02 $\pm$ 0.03 & 0.02 $\pm$ 0.03 & 0.02 $\pm$ 0.03 \\
 & Multivariate cox & 0.05 $\pm$ 0.04 & 0.07 $\pm$ 0.05 & 0.02 $\pm$ 0.03 & 0.03 $\pm$ 0.03 \\
 & Multivariate tree & 0.05 $\pm$ 0.04 & 0.05 $\pm$ 0.04 & 0.05 $\pm$ 0.04 & 0.05 $\pm$ 0.04 \\
 & Oracle & 0.06 $\pm$ 0.05 & 0.03 $\pm$ 0.03 & 0.03 $\pm$ 0.03 & 0.08 $\pm$ 0.05 \\
 & Univariate interaction & 0.02 $\pm$ 0.03 & 0.00 $\pm$ 0.00 & 0.00 $\pm$ 0.00 & 0.00 $\pm$ 0.00 \\
 & Univariate t-test & 0.06 $\pm$ 0.05 & 0.08 $\pm$ 0.05 & 0.08 $\pm$ 0.05 & 0.03 $\pm$ 0.03 \\
\cline{1-6}
\multirow[t]{8}{*}{0.145721} & ARDP & 0.05 $\pm$ 0.04 & 0.05 $\pm$ 0.04 & 0.03 $\pm$ 0.03 & 0.04 $\pm$ 0.04 \\
 & ITree & 0.00 $\pm$ 0.00 & 0.00 $\pm$ 0.00 & 0.00 $\pm$ 0.00 & 0.00 $\pm$ 0.00 \\
 & MOB & 0.04 $\pm$ 0.04 & 0.06 $\pm$ 0.05 & 0.03 $\pm$ 0.03 & 0.09 $\pm$ 0.06 \\
 & Multivariate cox & 0.03 $\pm$ 0.03 & 0.10 $\pm$ 0.06 & 0.03 $\pm$ 0.03 & 0.03 $\pm$ 0.03 \\
 & Multivariate tree & 0.07 $\pm$ 0.05 & 0.07 $\pm$ 0.05 & 0.07 $\pm$ 0.05 & 0.07 $\pm$ 0.05 \\
 & Oracle & 0.92 $\pm$ 0.05 & 0.90 $\pm$ 0.06 & 0.83 $\pm$ 0.07 & 0.57 $\pm$ 0.10 \\
 & Univariate interaction & 0.01 $\pm$ 0.02 & 0.00 $\pm$ 0.00 & 0.00 $\pm$ 0.00 & 0.00 $\pm$ 0.00 \\
 & Univariate t-test & 0.08 $\pm$ 0.05 & 0.02 $\pm$ 0.03 & 0.04 $\pm$ 0.04 & 0.03 $\pm$ 0.03 \\
\cline{1-6}
\multirow[t]{8}{*}{0.291443} & ARDP & 0.05 $\pm$ 0.04 & 0.03 $\pm$ 0.03 & 0.07 $\pm$ 0.05 & 0.03 $\pm$ 0.03 \\
 & ITree & 0.18 $\pm$ 0.08 & 0.13 $\pm$ 0.07 & 0.09 $\pm$ 0.06 & 0.01 $\pm$ 0.02 \\
 & MOB & 0.14 $\pm$ 0.07 & 0.11 $\pm$ 0.06 & 0.07 $\pm$ 0.05 & 0.09 $\pm$ 0.06 \\
 & Multivariate cox & 0.02 $\pm$ 0.03 & 0.04 $\pm$ 0.04 & 0.05 $\pm$ 0.04 & 0.00 $\pm$ 0.00 \\
 & Multivariate tree & 0.04 $\pm$ 0.04 & 0.04 $\pm$ 0.04 & 0.04 $\pm$ 0.04 & 0.04 $\pm$ 0.04 \\
 & Oracle & 1.00 $\pm$ 0.00 & 1.00 $\pm$ 0.00 & 1.00 $\pm$ 0.00 & 1.00 $\pm$ 0.00 \\
 & Univariate interaction & 0.10 $\pm$ 0.06 & 0.07 $\pm$ 0.05 & 0.03 $\pm$ 0.03 & 0.00 $\pm$ 0.00 \\
 & Univariate t-test & 0.04 $\pm$ 0.04 & 0.04 $\pm$ 0.04 & 0.03 $\pm$ 0.03 & 0.03 $\pm$ 0.03 \\
\cline{1-6}
\multirow[t]{8}{*}{0.437164} & ARDP & 0.07 $\pm$ 0.05 & 0.07 $\pm$ 0.05 & 0.10 $\pm$ 0.06 & 0.05 $\pm$ 0.04 \\
 & ITree & 0.83 $\pm$ 0.07 & 0.92 $\pm$ 0.05 & 0.82 $\pm$ 0.08 & 0.53 $\pm$ 0.10 \\
 & MOB & 0.69 $\pm$ 0.09 & 0.65 $\pm$ 0.09 & 0.48 $\pm$ 0.10 & 0.42 $\pm$ 0.10 \\
 & Multivariate cox & 0.03 $\pm$ 0.03 & 0.03 $\pm$ 0.03 & 0.02 $\pm$ 0.03 & 0.06 $\pm$ 0.05 \\
 & Multivariate tree & 0.51 $\pm$ 0.10 & 0.40 $\pm$ 0.10 & 0.25 $\pm$ 0.09 & 0.05 $\pm$ 0.04 \\
 & Oracle & 1.00 $\pm$ 0.00 & 1.00 $\pm$ 0.00 & 1.00 $\pm$ 0.00 & 1.00 $\pm$ 0.00 \\
 & Univariate interaction & 0.56 $\pm$ 0.10 & 0.83 $\pm$ 0.07 & 0.56 $\pm$ 0.10 & 0.37 $\pm$ 0.10 \\
 & Univariate t-test & 0.13 $\pm$ 0.07 & 0.14 $\pm$ 0.07 & 0.10 $\pm$ 0.06 & 0.08 $\pm$ 0.05 \\
\cline{1-6}
\hline
\end{tabular}
\end{center}
\end{table}

\newpage

\begin{table}[!htb]
\begin{center}
\caption{Averaged precision score of each method for p=20 for the subgroup definition with 4 predictive variables that are not prognostic.}
\label{sup:dim20_prog0_averaged_precision_score}
\begin{tabular}{llcccc}
\hline
arr & method &  0 & 1 & 2 & 3  \\
\hline
\multirow[t]{10}{*}{0.000000} & ARDP & 0.33 $\pm$ 0.03 & 0.29 $\pm$ 0.02 & 0.31 $\pm$ 0.03 & 0.32 $\pm$ 0.03 \\
 & ITree & 0.29 $\pm$ 0.02 & 0.30 $\pm$ 0.03 & 0.30 $\pm$ 0.02 & 0.28 $\pm$ 0.02 \\
 & MOB & 0.32 $\pm$ 0.03 & 0.31 $\pm$ 0.03 & 0.32 $\pm$ 0.03 & 0.30 $\pm$ 0.02 \\
 & Multivariate cox & 0.20 $\pm$ 0.01 & 0.18 $\pm$ 0.01 & 0.19 $\pm$ 0.01 & 0.18 $\pm$ 0.01 \\
 & Multivariate tree & 0.32 $\pm$ 0.03 & 0.29 $\pm$ 0.03 & 0.32 $\pm$ 0.03 & 0.31 $\pm$ 0.03 \\
 & Oracle & 0.30 $\pm$ 0.03 & 0.30 $\pm$ 0.02 & 0.31 $\pm$ 0.02 & 0.31 $\pm$ 0.03 \\
 & SIDES & 0.31 $\pm$ 0.02 & 0.31 $\pm$ 0.03 & 0.30 $\pm$ 0.03 & 0.29 $\pm$ 0.02 \\
 & SeqBT & 0.23 $\pm$ 0.01 & 0.25 $\pm$ 0.02 & 0.25 $\pm$ 0.02 & 0.26 $\pm$ 0.02 \\
 & Univariate interaction & 0.28 $\pm$ 0.02 & 0.27 $\pm$ 0.02 & 0.26 $\pm$ 0.02 & 0.24 $\pm$ 0.02 \\
 & Univariate t-test & 0.30 $\pm$ 0.02 & 0.31 $\pm$ 0.03 & 0.31 $\pm$ 0.02 & 0.34 $\pm$ 0.03 \\
\cline{1-6}
\multirow[t]{10}{*}{0.144137} & ARDP & 0.31 $\pm$ 0.02 & 0.29 $\pm$ 0.02 & 0.32 $\pm$ 0.03 & 0.32 $\pm$ 0.03 \\
 & ITree & 0.40 $\pm$ 0.03 & 0.42 $\pm$ 0.03 & 0.42 $\pm$ 0.03 & 0.40 $\pm$ 0.03 \\
 & MOB & 0.37 $\pm$ 0.03 & 0.40 $\pm$ 0.03 & 0.37 $\pm$ 0.03 & 0.34 $\pm$ 0.03 \\
 & Multivariate cox & 0.56 $\pm$ 0.04 & 0.51 $\pm$ 0.03 & 0.43 $\pm$ 0.03 & 0.33 $\pm$ 0.03 \\
 & Multivariate tree & 0.27 $\pm$ 0.02 & 0.32 $\pm$ 0.03 & 0.32 $\pm$ 0.03 & 0.32 $\pm$ 0.03 \\
 & Oracle & 0.30 $\pm$ 0.02 & 0.29 $\pm$ 0.02 & 0.30 $\pm$ 0.02 & 0.31 $\pm$ 0.03 \\
 & SIDES & 0.33 $\pm$ 0.03 & 0.31 $\pm$ 0.03 & 0.32 $\pm$ 0.03 & 0.31 $\pm$ 0.02 \\
 & SeqBT & 0.29 $\pm$ 0.02 & 0.29 $\pm$ 0.02 & 0.28 $\pm$ 0.02 & 0.29 $\pm$ 0.02 \\
 & Univariate interaction & 0.49 $\pm$ 0.04 & 0.49 $\pm$ 0.04 & 0.44 $\pm$ 0.04 & 0.40 $\pm$ 0.04 \\
 & Univariate t-test & 0.40 $\pm$ 0.03 & 0.37 $\pm$ 0.03 & 0.35 $\pm$ 0.03 & 0.35 $\pm$ 0.03 \\
\cline{1-6}
\multirow[t]{10}{*}{0.288274} & ARDP & 0.31 $\pm$ 0.03 & 0.28 $\pm$ 0.02 & 0.30 $\pm$ 0.03 & 0.34 $\pm$ 0.04 \\
 & ITree & 0.58 $\pm$ 0.03 & 0.58 $\pm$ 0.03 & 0.55 $\pm$ 0.03 & 0.47 $\pm$ 0.02 \\
 & MOB & 0.55 $\pm$ 0.04 & 0.55 $\pm$ 0.03 & 0.54 $\pm$ 0.03 & 0.47 $\pm$ 0.03 \\
 & Multivariate cox & 0.84 $\pm$ 0.03 & 0.83 $\pm$ 0.03 & 0.76 $\pm$ 0.03 & 0.63 $\pm$ 0.03 \\
 & Multivariate tree & 0.32 $\pm$ 0.03 & 0.29 $\pm$ 0.03 & 0.31 $\pm$ 0.02 & 0.33 $\pm$ 0.03 \\
 & Oracle & 0.33 $\pm$ 0.03 & 0.29 $\pm$ 0.02 & 0.31 $\pm$ 0.03 & 0.30 $\pm$ 0.03 \\
 & SIDES & 0.38 $\pm$ 0.03 & 0.35 $\pm$ 0.03 & 0.33 $\pm$ 0.03 & 0.32 $\pm$ 0.03 \\
 & SeqBT & 0.49 $\pm$ 0.05 & 0.46 $\pm$ 0.04 & 0.42 $\pm$ 0.05 & 0.38 $\pm$ 0.03 \\
 & Univariate interaction & 0.79 $\pm$ 0.03 & 0.79 $\pm$ 0.03 & 0.76 $\pm$ 0.03 & 0.67 $\pm$ 0.04 \\
 & Univariate t-test & 0.46 $\pm$ 0.03 & 0.43 $\pm$ 0.04 & 0.43 $\pm$ 0.03 & 0.38 $\pm$ 0.03 \\
\cline{1-6}
\multirow[t]{10}{*}{0.432411} & ARDP & 0.26 $\pm$ 0.02 & 0.28 $\pm$ 0.02 & 0.30 $\pm$ 0.03 & 0.29 $\pm$ 0.03 \\
 & ITree & 0.76 $\pm$ 0.02 & 0.83 $\pm$ 0.02 & 0.85 $\pm$ 0.02 & 0.78 $\pm$ 0.02 \\
 & MOB & 0.77 $\pm$ 0.03 & 0.74 $\pm$ 0.03 & 0.74 $\pm$ 0.03 & 0.71 $\pm$ 0.03 \\
 & Multivariate cox & 0.86 $\pm$ 0.03 & 0.88 $\pm$ 0.02 & 0.88 $\pm$ 0.02 & 0.83 $\pm$ 0.03 \\
 & Multivariate tree & 0.28 $\pm$ 0.03 & 0.31 $\pm$ 0.02 & 0.30 $\pm$ 0.02 & 0.31 $\pm$ 0.03 \\
 & Oracle & 0.30 $\pm$ 0.03 & 0.31 $\pm$ 0.03 & 0.29 $\pm$ 0.02 & 0.31 $\pm$ 0.03 \\
 & SIDES & 0.37 $\pm$ 0.03 & 0.34 $\pm$ 0.03 & 0.33 $\pm$ 0.03 & 0.33 $\pm$ 0.03 \\
 & SeqBT & 0.76 $\pm$ 0.03 & 0.78 $\pm$ 0.03 & 0.81 $\pm$ 0.04 & 0.71 $\pm$ 0.05 \\
 & Univariate interaction & 0.98 $\pm$ 0.01 & 0.99 $\pm$ 0.01 & 0.98 $\pm$ 0.01 & 0.96 $\pm$ 0.01 \\
 & Univariate t-test & 0.42 $\pm$ 0.03 & 0.41 $\pm$ 0.03 & 0.40 $\pm$ 0.03 & 0.39 $\pm$ 0.03 \\
\cline{1-6}
\hline
\end{tabular}
\end{center}
\end{table}

\begin{table}[!htb]
\begin{center}
\caption{Averaged precision score of each method for p=100 for the subgroup definition with 4 predictive variables that are not prognostic.}
\label{sup:dim100_prog0_averaged_precision_score}
\begin{tabular}{llcccc}
\hline
arr & method & 0 & 1 & 2 & 3  \\
\hline
\multirow[t]{8}{*}{0.000000} & ARDP & 0.08 $\pm$ 0.01 & 0.08 $\pm$ 0.01 & 0.07 $\pm$ 0.01 & 0.07 $\pm$ 0.01 \\
 & ITree & 0.08 $\pm$ 0.02 & 0.08 $\pm$ 0.01 & 0.08 $\pm$ 0.02 & 0.08 $\pm$ 0.01 \\
 & MOB & 0.07 $\pm$ 0.01 & 0.07 $\pm$ 0.01 & 0.08 $\pm$ 0.01 & 0.09 $\pm$ 0.02 \\
 & Multivariate cox & 0.06 $\pm$ 0.01 & 0.06 $\pm$ 0.01 & 0.06 $\pm$ 0.01 & 0.06 $\pm$ 0.01 \\
 & Multivariate tree & 0.07 $\pm$ 0.01 & 0.09 $\pm$ 0.02 & 0.07 $\pm$ 0.01 & 0.07 $\pm$ 0.01 \\
 & Oracle & 0.09 $\pm$ 0.02 & 0.09 $\pm$ 0.01 & 0.08 $\pm$ 0.01 & 0.09 $\pm$ 0.02 \\
 & Univariate interaction & 0.08 $\pm$ 0.01 & 0.08 $\pm$ 0.01 & 0.08 $\pm$ 0.01 & 0.08 $\pm$ 0.01 \\
 & Univariate t-test & 0.09 $\pm$ 0.02 & 0.09 $\pm$ 0.01 & 0.08 $\pm$ 0.02 & 0.08 $\pm$ 0.01 \\
\cline{1-6}
\multirow[t]{8}{*}{0.144367} & ARDP & 0.08 $\pm$ 0.01 & 0.08 $\pm$ 0.01 & 0.07 $\pm$ 0.01 & 0.07 $\pm$ 0.01 \\
 & ITree & 0.16 $\pm$ 0.02 & 0.16 $\pm$ 0.03 & 0.14 $\pm$ 0.03 & 0.11 $\pm$ 0.02 \\
 & MOB & 0.12 $\pm$ 0.02 & 0.12 $\pm$ 0.02 & 0.11 $\pm$ 0.02 & 0.10 $\pm$ 0.02 \\
 & Multivariate cox & 0.25 $\pm$ 0.03 & 0.17 $\pm$ 0.02 & 0.13 $\pm$ 0.02 & 0.10 $\pm$ 0.01 \\
 & Multivariate tree & 0.07 $\pm$ 0.01 & 0.07 $\pm$ 0.01 & 0.08 $\pm$ 0.02 & 0.08 $\pm$ 0.01 \\
 & Oracle & 0.08 $\pm$ 0.01 & 0.09 $\pm$ 0.01 & 0.10 $\pm$ 0.02 & 0.08 $\pm$ 0.01 \\
 & Univariate interaction & 0.20 $\pm$ 0.03 & 0.19 $\pm$ 0.03 & 0.16 $\pm$ 0.03 & 0.13 $\pm$ 0.02 \\
 & Univariate t-test & 0.10 $\pm$ 0.01 & 0.10 $\pm$ 0.02 & 0.10 $\pm$ 0.02 & 0.08 $\pm$ 0.01 \\
\cline{1-6}
\multirow[t]{8}{*}{0.288733} & ARDP & 0.07 $\pm$ 0.01 & 0.08 $\pm$ 0.01 & 0.08 $\pm$ 0.01 & 0.07 $\pm$ 0.01 \\
 & ITree & 0.33 $\pm$ 0.03 & 0.35 $\pm$ 0.03 & 0.28 $\pm$ 0.02 & 0.24 $\pm$ 0.03 \\
 & MOB & 0.27 $\pm$ 0.03 & 0.26 $\pm$ 0.03 & 0.22 $\pm$ 0.03 & 0.16 $\pm$ 0.02 \\
 & Multivariate cox & 0.47 $\pm$ 0.04 & 0.43 $\pm$ 0.04 & 0.36 $\pm$ 0.04 & 0.26 $\pm$ 0.03 \\
 & Multivariate tree & 0.08 $\pm$ 0.01 & 0.09 $\pm$ 0.01 & 0.08 $\pm$ 0.02 & 0.08 $\pm$ 0.01 \\
 & Oracle & 0.07 $\pm$ 0.01 & 0.07 $\pm$ 0.01 & 0.08 $\pm$ 0.01 & 0.08 $\pm$ 0.01 \\
 & Univariate interaction & 0.59 $\pm$ 0.04 & 0.60 $\pm$ 0.04 & 0.51 $\pm$ 0.04 & 0.42 $\pm$ 0.04 \\
 & Univariate t-test & 0.20 $\pm$ 0.03 & 0.19 $\pm$ 0.03 & 0.14 $\pm$ 0.02 & 0.13 $\pm$ 0.02 \\
\cline{1-6}
\multirow[t]{8}{*}{0.433100} & ARDP & 0.07 $\pm$ 0.01 & 0.07 $\pm$ 0.01 & 0.07 $\pm$ 0.01 & 0.07 $\pm$ 0.01 \\
 & ITree & 0.54 $\pm$ 0.04 & 0.64 $\pm$ 0.03 & 0.64 $\pm$ 0.04 & 0.58 $\pm$ 0.04 \\
 & MOB & 0.49 $\pm$ 0.03 & 0.52 $\pm$ 0.04 & 0.48 $\pm$ 0.03 & 0.44 $\pm$ 0.03 \\
 & Multivariate cox & 0.58 $\pm$ 0.04 & 0.63 $\pm$ 0.04 & 0.57 $\pm$ 0.04 & 0.50 $\pm$ 0.04 \\
 & Multivariate tree & 0.08 $\pm$ 0.01 & 0.09 $\pm$ 0.01 & 0.08 $\pm$ 0.01 & 0.08 $\pm$ 0.01 \\
 & Oracle & 0.08 $\pm$ 0.01 & 0.07 $\pm$ 0.01 & 0.08 $\pm$ 0.01 & 0.08 $\pm$ 0.01 \\
 & Univariate interaction & 0.91 $\pm$ 0.02 & 0.95 $\pm$ 0.02 & 0.94 $\pm$ 0.02 & 0.92 $\pm$ 0.02 \\
 & Univariate t-test & 0.27 $\pm$ 0.03 & 0.26 $\pm$ 0.03 & 0.23 $\pm$ 0.03 & 0.20 $\pm$ 0.03 \\
\cline{1-6}
\hline
\end{tabular}
\end{center}
\end{table}

\begin{table}[!htb]
\begin{center}
\caption{Averaged precision score of each method for p=1000 for the subgroup definition with 4 predictive variables that are not prognostic.}
\label{sup:dim1000_prog0_averaged_precision_score}
\begin{tabular}{llcccc}
\hline
arr & method & 0 & 1 & 2 & 3 \\
\hline
\multirow[t]{8}{*}{0.000000} & ARDP & 0.01 $\pm$ 0.00 & 0.01 $\pm$ 0.00 & 0.01 $\pm$ 0.01 & 0.01 $\pm$ 0.01 \\
 & ITree & 0.01 $\pm$ 0.00 & 0.01 $\pm$ 0.00 & 0.01 $\pm$ 0.00 & 0.01 $\pm$ 0.01 \\
 & MOB & 0.01 $\pm$ 0.00 & 0.01 $\pm$ 0.01 & 0.01 $\pm$ 0.01 & 0.01 $\pm$ 0.01 \\
 & Multivariate cox & 0.01 $\pm$ 0.00 & 0.01 $\pm$ 0.01 & 0.01 $\pm$ 0.00 & 0.01 $\pm$ 0.00 \\
 & Multivariate tree & 0.01 $\pm$ 0.00 & 0.01 $\pm$ 0.00 & 0.01 $\pm$ 0.00 & 0.01 $\pm$ 0.00 \\
 & Oracle & 0.01 $\pm$ 0.00 & 0.01 $\pm$ 0.00 & 0.01 $\pm$ 0.00 & 0.01 $\pm$ 0.00 \\
 & Univariate interaction & 0.01 $\pm$ 0.00 & 0.01 $\pm$ 0.00 & 0.01 $\pm$ 0.01 & 0.01 $\pm$ 0.01 \\
 & Univariate t-test & 0.01 $\pm$ 0.00 & 0.01 $\pm$ 0.00 & 0.01 $\pm$ 0.00 & 0.01 $\pm$ 0.00 \\
\cline{1-6}
\multirow[t]{8}{*}{0.145721} & ARDP & 0.01 $\pm$ 0.00 & 0.01 $\pm$ 0.00 & 0.01 $\pm$ 0.00 & 0.01 $\pm$ 0.00 \\
 & ITree & 0.03 $\pm$ 0.01 & 0.02 $\pm$ 0.01 & 0.03 $\pm$ 0.01 & 0.02 $\pm$ 0.01 \\
 & MOB & 0.02 $\pm$ 0.01 & 0.02 $\pm$ 0.01 & 0.01 $\pm$ 0.01 & 0.01 $\pm$ 0.01 \\
 & Multivariate cox & 0.02 $\pm$ 0.01 & 0.02 $\pm$ 0.01 & 0.02 $\pm$ 0.01 & 0.02 $\pm$ 0.01 \\
 & Multivariate tree & 0.01 $\pm$ 0.00 & 0.01 $\pm$ 0.00 & 0.01 $\pm$ 0.01 & 0.01 $\pm$ 0.01 \\
 & Oracle & 0.01 $\pm$ 0.00 & 0.01 $\pm$ 0.01 & 0.01 $\pm$ 0.00 & 0.01 $\pm$ 0.00 \\
 & Univariate interaction & 0.04 $\pm$ 0.02 & 0.04 $\pm$ 0.02 & 0.04 $\pm$ 0.01 & 0.03 $\pm$ 0.01 \\
 & Univariate t-test & 0.01 $\pm$ 0.00 & 0.01 $\pm$ 0.00 & 0.02 $\pm$ 0.01 & 0.01 $\pm$ 0.01 \\
\cline{1-6}
\multirow[t]{8}{*}{0.291443} & ARDP & 0.01 $\pm$ 0.00 & 0.01 $\pm$ 0.00 & 0.01 $\pm$ 0.00 & 0.01 $\pm$ 0.00 \\
 & ITree & 0.18 $\pm$ 0.03 & 0.17 $\pm$ 0.02 & 0.16 $\pm$ 0.03 & 0.13 $\pm$ 0.02 \\
 & MOB & 0.08 $\pm$ 0.02 & 0.09 $\pm$ 0.02 & 0.08 $\pm$ 0.02 & 0.08 $\pm$ 0.02 \\
 & Multivariate cox & 0.15 $\pm$ 0.03 & 0.14 $\pm$ 0.03 & 0.12 $\pm$ 0.03 & 0.07 $\pm$ 0.02 \\
 & Multivariate tree & 0.01 $\pm$ 0.00 & 0.01 $\pm$ 0.00 & 0.01 $\pm$ 0.00 & 0.01 $\pm$ 0.01 \\
 & Oracle & 0.01 $\pm$ 0.00 & 0.01 $\pm$ 0.00 & 0.01 $\pm$ 0.01 & 0.01 $\pm$ 0.00 \\
 & Univariate interaction & 0.27 $\pm$ 0.04 & 0.30 $\pm$ 0.04 & 0.25 $\pm$ 0.04 & 0.19 $\pm$ 0.03 \\
 & Univariate t-test & 0.04 $\pm$ 0.01 & 0.03 $\pm$ 0.01 & 0.02 $\pm$ 0.01 & 0.02 $\pm$ 0.01 \\
\cline{1-6}
\multirow[t]{8}{*}{0.437164} & ARDP & 0.01 $\pm$ 0.01 & 0.01 $\pm$ 0.01 & 0.01 $\pm$ 0.00 & 0.01 $\pm$ 0.00 \\
 & ITree & 0.32 $\pm$ 0.03 & 0.41 $\pm$ 0.03 & 0.37 $\pm$ 0.03 & 0.32 $\pm$ 0.03 \\
 & MOB & 0.30 $\pm$ 0.03 & 0.30 $\pm$ 0.03 & 0.26 $\pm$ 0.03 & 0.22 $\pm$ 0.03 \\
 & Multivariate cox & 0.35 $\pm$ 0.04 & 0.34 $\pm$ 0.05 & 0.27 $\pm$ 0.04 & 0.19 $\pm$ 0.03 \\
 & Multivariate tree & 0.01 $\pm$ 0.01 & 0.01 $\pm$ 0.01 & 0.01 $\pm$ 0.00 & 0.01 $\pm$ 0.00 \\
 & Oracle & 0.01 $\pm$ 0.01 & 0.01 $\pm$ 0.01 & 0.01 $\pm$ 0.00 & 0.01 $\pm$ 0.01 \\
 & Univariate interaction & 0.72 $\pm$ 0.04 & 0.84 $\pm$ 0.03 & 0.78 $\pm$ 0.04 & 0.68 $\pm$ 0.04 \\
 & Univariate t-test & 0.14 $\pm$ 0.03 & 0.12 $\pm$ 0.02 & 0.09 $\pm$ 0.02 & 0.05 $\pm$ 0.02 \\
\cline{1-6}
\hline
\end{tabular}
\end{center}
\end{table}

\newpage

\begin{table}[!htb]
\begin{center}
\caption{Accuracy of each method for p=20 for the subgroup definition with 4 predictive variables that are not prognostic.}
\label{sup:dim20_prog0_accuracy}
\begin{tabular}{llcccc}
\hline
arr & method & 0 & 1 & 2 & 3  \\
\hline
\multirow[t]{10}{*}{0.000000} & ARDP & 0.51 $\pm$ 0.01 & 0.51 $\pm$ 0.01 & 0.52 $\pm$ 0.01 & 0.52 $\pm$ 0.01 \\
 & ITree & 0.50 $\pm$ 0.00 & 0.50 $\pm$ 0.00 & 0.50 $\pm$ 0.00 & 0.50 $\pm$ 0.00 \\
 & MOB & 0.50 $\pm$ 0.00 & 0.50 $\pm$ 0.00 & 0.50 $\pm$ 0.00 & 0.50 $\pm$ 0.00 \\
 & Multivariate cox & 0.50 $\pm$ 0.01 & 0.50 $\pm$ 0.01 & 0.50 $\pm$ 0.01 & 0.50 $\pm$ 0.01 \\
 & Multivariate tree & 0.50 $\pm$ 0.00 & 0.50 $\pm$ 0.00 & 0.50 $\pm$ 0.00 & 0.50 $\pm$ 0.00 \\
 & SIDES & 0.50 $\pm$ 0.00 & 0.50 $\pm$ 0.00 & 0.50 $\pm$ 0.00 & 0.50 $\pm$ 0.00 \\
 & SeqBT & 0.50 $\pm$ 0.00 & 0.50 $\pm$ 0.00 & 0.51 $\pm$ 0.01 & 0.51 $\pm$ 0.00 \\
 & Univariate interaction & 0.51 $\pm$ 0.01 & 0.52 $\pm$ 0.01 & 0.51 $\pm$ 0.01 & 0.51 $\pm$ 0.00 \\
 & Univariate t-test & 0.51 $\pm$ 0.01 & 0.52 $\pm$ 0.01 & 0.52 $\pm$ 0.01 & 0.53 $\pm$ 0.01 \\
\cline{1-6}
\multirow[t]{10}{*}{0.144137} & ARDP & 0.49 $\pm$ 0.00 & 0.49 $\pm$ 0.00 & 0.50 $\pm$ 0.00 & 0.50 $\pm$ 0.01 \\
 & ITree & 0.51 $\pm$ 0.01 & 0.51 $\pm$ 0.01 & 0.50 $\pm$ 0.01 & 0.50 $\pm$ 0.00 \\
 & MOB & 0.50 $\pm$ 0.00 & 0.50 $\pm$ 0.00 & 0.50 $\pm$ 0.00 & 0.51 $\pm$ 0.00 \\
 & Multivariate cox & 0.64 $\pm$ 0.01 & 0.63 $\pm$ 0.01 & 0.61 $\pm$ 0.01 & 0.59 $\pm$ 0.01 \\
 & Multivariate tree & 0.50 $\pm$ 0.00 & 0.50 $\pm$ 0.00 & 0.50 $\pm$ 0.00 & 0.50 $\pm$ 0.00 \\
 & SIDES & 0.50 $\pm$ 0.00 & 0.50 $\pm$ 0.00 & 0.50 $\pm$ 0.00 & 0.50 $\pm$ 0.00 \\
 & SeqBT & 0.52 $\pm$ 0.01 & 0.52 $\pm$ 0.01 & 0.52 $\pm$ 0.01 & 0.52 $\pm$ 0.01 \\
 & Univariate interaction & 0.55 $\pm$ 0.01 & 0.55 $\pm$ 0.01 & 0.54 $\pm$ 0.01 & 0.53 $\pm$ 0.01 \\
 & Univariate t-test & 0.54 $\pm$ 0.01 & 0.53 $\pm$ 0.01 & 0.53 $\pm$ 0.01 & 0.53 $\pm$ 0.01 \\
\cline{1-6}
\multirow[t]{10}{*}{0.288274} & ARDP & 0.49 $\pm$ 0.00 & 0.49 $\pm$ 0.00 & 0.49 $\pm$ 0.00 & 0.49 $\pm$ 0.00 \\
 & ITree & 0.65 $\pm$ 0.02 & 0.63 $\pm$ 0.02 & 0.59 $\pm$ 0.02 & 0.54 $\pm$ 0.01 \\
 & MOB & 0.52 $\pm$ 0.01 & 0.52 $\pm$ 0.01 & 0.51 $\pm$ 0.01 & 0.51 $\pm$ 0.00 \\
 & Multivariate cox & 0.68 $\pm$ 0.01 & 0.68 $\pm$ 0.01 & 0.66 $\pm$ 0.01 & 0.64 $\pm$ 0.01 \\
 & Multivariate tree & 0.50 $\pm$ 0.00 & 0.50 $\pm$ 0.00 & 0.50 $\pm$ 0.00 & 0.50 $\pm$ 0.00 \\
 & SIDES & 0.49 $\pm$ 0.00 & 0.49 $\pm$ 0.00 & 0.49 $\pm$ 0.00 & 0.49 $\pm$ 0.00 \\
 & SeqBT & 0.61 $\pm$ 0.02 & 0.61 $\pm$ 0.02 & 0.59 $\pm$ 0.02 & 0.56 $\pm$ 0.02 \\
 & Univariate interaction & 0.58 $\pm$ 0.01 & 0.58 $\pm$ 0.01 & 0.59 $\pm$ 0.01 & 0.57 $\pm$ 0.01 \\
 & Univariate t-test & 0.54 $\pm$ 0.01 & 0.54 $\pm$ 0.01 & 0.54 $\pm$ 0.01 & 0.53 $\pm$ 0.01 \\
\cline{1-6}
\multirow[t]{10}{*}{0.432411} & ARDP & 0.50 $\pm$ 0.00 & 0.50 $\pm$ 0.00 & 0.50 $\pm$ 0.00 & 0.50 $\pm$ 0.00 \\
 & ITree & 0.74 $\pm$ 0.01 & 0.77 $\pm$ 0.01 & 0.77 $\pm$ 0.01 & 0.75 $\pm$ 0.01 \\
 & MOB & 0.54 $\pm$ 0.00 & 0.54 $\pm$ 0.00 & 0.54 $\pm$ 0.00 & 0.53 $\pm$ 0.00 \\
 & Multivariate cox & 0.68 $\pm$ 0.01 & 0.69 $\pm$ 0.00 & 0.68 $\pm$ 0.01 & 0.66 $\pm$ 0.01 \\
 & Multivariate tree & 0.72 $\pm$ 0.03 & 0.69 $\pm$ 0.04 & 0.66 $\pm$ 0.03 & 0.60 $\pm$ 0.03 \\
 & SIDES & 0.49 $\pm$ 0.00 & 0.50 $\pm$ 0.00 & 0.50 $\pm$ 0.00 & 0.50 $\pm$ 0.00 \\
 & SeqBT & 0.75 $\pm$ 0.01 & 0.75 $\pm$ 0.01 & 0.76 $\pm$ 0.02 & 0.71 $\pm$ 0.02 \\
 & Univariate interaction & 0.59 $\pm$ 0.00 & 0.59 $\pm$ 0.00 & 0.59 $\pm$ 0.00 & 0.59 $\pm$ 0.00 \\
 & Univariate t-test & 0.53 $\pm$ 0.01 & 0.53 $\pm$ 0.01 & 0.53 $\pm$ 0.01 & 0.53 $\pm$ 0.01 \\
\cline{1-6}
\hline
\end{tabular}
\end{center}
\end{table}

\newpage 

\begin{table}[!htb]
\begin{center}
\caption{Accuracy of each method for p=100 for the subgroup definition with 4 predictive variables that are not prognostic.}
\label{sup:dim100_prog0_accuracy}
\begin{tabular}{llcccc}
\hline
arr & method & 0 & 1 & 2 & 3 \\
\hline
\multirow[t]{8}{*}{0.000000} & ARDP & 0.50 $\pm$ 0.00 & 0.50 $\pm$ 0.00 & 0.50 $\pm$ 0.00 & 0.50 $\pm$ 0.00 \\
 & ITree & 0.50 $\pm$ 0.00 & 0.50 $\pm$ 0.00 & 0.50 $\pm$ 0.00 & 0.50 $\pm$ 0.00 \\
 & MOB & 0.50 $\pm$ 0.00 & 0.50 $\pm$ 0.00 & 0.50 $\pm$ 0.00 & 0.50 $\pm$ 0.00 \\
 & Multivariate cox & 0.50 $\pm$ 0.01 & 0.50 $\pm$ 0.01 & 0.50 $\pm$ 0.01 & 0.50 $\pm$ 0.01 \\
 & Multivariate tree & 0.50 $\pm$ 0.00 & 0.50 $\pm$ 0.00 & 0.50 $\pm$ 0.00 & 0.50 $\pm$ 0.00 \\
 & Univariate interaction & 0.50 $\pm$ 0.00 & 0.50 $\pm$ 0.00 & 0.50 $\pm$ 0.00 & 0.50 $\pm$ 0.00 \\
 & Univariate t-test & 0.51 $\pm$ 0.01 & 0.50 $\pm$ 0.00 & 0.50 $\pm$ 0.00 & 0.50 $\pm$ 0.00 \\
\cline{1-6}
\multirow[t]{8}{*}{0.144367} & ARDP & 0.50 $\pm$ 0.00 & 0.50 $\pm$ 0.00 & 0.50 $\pm$ 0.00 & 0.50 $\pm$ 0.00 \\
 & ITree & 0.50 $\pm$ 0.01 & 0.50 $\pm$ 0.00 & 0.50 $\pm$ 0.00 & 0.50 $\pm$ 0.00 \\
 & MOB & 0.50 $\pm$ 0.00 & 0.50 $\pm$ 0.00 & 0.50 $\pm$ 0.00 & 0.50 $\pm$ 0.00 \\
 & Multivariate cox & 0.57 $\pm$ 0.00 & 0.55 $\pm$ 0.00 & 0.55 $\pm$ 0.00 & 0.54 $\pm$ 0.00 \\
 & Multivariate tree & 0.50 $\pm$ 0.00 & 0.50 $\pm$ 0.00 & 0.50 $\pm$ 0.00 & 0.50 $\pm$ 0.00 \\
 & Univariate interaction & 0.53 $\pm$ 0.01 & 0.52 $\pm$ 0.01 & 0.51 $\pm$ 0.01 & 0.51 $\pm$ 0.01 \\
 & Univariate t-test & 0.51 $\pm$ 0.00 & 0.51 $\pm$ 0.01 & 0.51 $\pm$ 0.01 & 0.50 $\pm$ 0.00 \\
\cline{1-6}
\multirow[t]{8}{*}{0.288733} & ARDP & 0.50 $\pm$ 0.00 & 0.50 $\pm$ 0.00 & 0.50 $\pm$ 0.00 & 0.50 $\pm$ 0.00 \\
 & ITree & 0.59 $\pm$ 0.02 & 0.60 $\pm$ 0.02 & 0.55 $\pm$ 0.02 & 0.51 $\pm$ 0.01 \\
 & MOB & 0.52 $\pm$ 0.01 & 0.52 $\pm$ 0.01 & 0.51 $\pm$ 0.01 & 0.50 $\pm$ 0.00 \\
 & Multivariate cox & 0.59 $\pm$ 0.00 & 0.59 $\pm$ 0.00 & 0.58 $\pm$ 0.00 & 0.57 $\pm$ 0.00 \\
 & Multivariate tree & 0.50 $\pm$ 0.00 & 0.50 $\pm$ 0.00 & 0.50 $\pm$ 0.00 & 0.50 $\pm$ 0.00 \\
 & Univariate interaction & 0.58 $\pm$ 0.01 & 0.58 $\pm$ 0.01 & 0.57 $\pm$ 0.01 & 0.56 $\pm$ 0.01 \\
 & Univariate t-test & 0.52 $\pm$ 0.01 & 0.53 $\pm$ 0.01 & 0.51 $\pm$ 0.01 & 0.51 $\pm$ 0.01 \\
\cline{1-6}
\multirow[t]{8}{*}{0.433100} & ARDP & 0.50 $\pm$ 0.00 & 0.50 $\pm$ 0.00 & 0.50 $\pm$ 0.00 & 0.50 $\pm$ 0.00 \\
 & ITree & 0.72 $\pm$ 0.01 & 0.75 $\pm$ 0.01 & 0.75 $\pm$ 0.01 & 0.72 $\pm$ 0.01 \\
 & MOB & 0.54 $\pm$ 0.00 & 0.53 $\pm$ 0.00 & 0.54 $\pm$ 0.00 & 0.53 $\pm$ 0.01 \\
 & Multivariate cox & 0.60 $\pm$ 0.00 & 0.60 $\pm$ 0.00 & 0.60 $\pm$ 0.00 & 0.59 $\pm$ 0.00 \\
 & Multivariate tree & 0.69 $\pm$ 0.03 & 0.67 $\pm$ 0.03 & 0.63 $\pm$ 0.03 & 0.57 $\pm$ 0.03 \\
 & Univariate interaction & 0.59 $\pm$ 0.00 & 0.59 $\pm$ 0.00 & 0.60 $\pm$ 0.00 & 0.60 $\pm$ 0.00 \\
 & Univariate t-test & 0.53 $\pm$ 0.01 & 0.52 $\pm$ 0.01 & 0.53 $\pm$ 0.01 & 0.52 $\pm$ 0.01 \\
\cline{1-6}
\hline
\end{tabular}
\end{center}
\end{table}

\newpage 

\begin{table}[!htb]
\begin{center}
\caption{Accuracy of each method for p=1000 for the subgroup definition with 4 predictive variables that are not prognostic.}
\label{sup:dim1000_prog0_accuracy}
\begin{tabular}{llcccc}
\hline
arr & method & 0 & 1 & 2 & 3 \\
\hline
\multirow[t]{8}{*}{0.000000} & ARDP & 0.50 $\pm$ 0.00 & 0.50 $\pm$ 0.00 & 0.50 $\pm$ 0.00 & 0.50 $\pm$ 0.00 \\
 & ITree & 0.50 $\pm$ 0.00 & 0.50 $\pm$ 0.00 & 0.50 $\pm$ 0.00 & 0.50 $\pm$ 0.00 \\
 & MOB & 0.50 $\pm$ 0.00 & 0.50 $\pm$ 0.00 & 0.50 $\pm$ 0.00 & 0.50 $\pm$ 0.00 \\
 & Multivariate cox & 0.50 $\pm$ 0.00 & 0.50 $\pm$ 0.00 & 0.50 $\pm$ 0.00 & 0.50 $\pm$ 0.00 \\
 & Multivariate tree & 0.50 $\pm$ 0.00 & 0.50 $\pm$ 0.00 & 0.50 $\pm$ 0.00 & 0.50 $\pm$ 0.00 \\
 & Univariate interaction & 0.50 $\pm$ 0.00 & 0.50 $\pm$ 0.00 & 0.50 $\pm$ 0.00 & 0.50 $\pm$ 0.00 \\
 & Univariate t-test & 0.50 $\pm$ 0.00 & 0.50 $\pm$ 0.00 & 0.50 $\pm$ 0.00 & 0.50 $\pm$ 0.00 \\
\cline{1-6}
\multirow[t]{8}{*}{0.145721} & ARDP & 0.50 $\pm$ 0.00 & 0.50 $\pm$ 0.00 & 0.50 $\pm$ 0.00 & 0.50 $\pm$ 0.00 \\
 & ITree & 0.50 $\pm$ 0.00 & 0.50 $\pm$ 0.00 & 0.50 $\pm$ 0.00 & 0.50 $\pm$ 0.00 \\
 & MOB & 0.50 $\pm$ 0.00 & 0.50 $\pm$ 0.00 & 0.50 $\pm$ 0.00 & 0.50 $\pm$ 0.00 \\
 & Multivariate cox & 0.52 $\pm$ 0.00 & 0.52 $\pm$ 0.00 & 0.51 $\pm$ 0.00 & 0.51 $\pm$ 0.00 \\
 & Multivariate tree & 0.50 $\pm$ 0.00 & 0.50 $\pm$ 0.00 & 0.50 $\pm$ 0.00 & 0.50 $\pm$ 0.00 \\
 & Univariate interaction & 0.50 $\pm$ 0.00 & 0.50 $\pm$ 0.01 & 0.50 $\pm$ 0.00 & 0.50 $\pm$ 0.00 \\
 & Univariate t-test & 0.50 $\pm$ 0.00 & 0.50 $\pm$ 0.00 & 0.50 $\pm$ 0.00 & 0.50 $\pm$ 0.00 \\
\cline{1-6}
\multirow[t]{8}{*}{0.291443} & ARDP & 0.50 $\pm$ 0.00 & 0.50 $\pm$ 0.00 & 0.50 $\pm$ 0.00 & 0.50 $\pm$ 0.00 \\
 & ITree & 0.53 $\pm$ 0.01 & 0.52 $\pm$ 0.01 & 0.52 $\pm$ 0.01 & 0.50 $\pm$ 0.00 \\
 & MOB & 0.50 $\pm$ 0.00 & 0.50 $\pm$ 0.00 & 0.50 $\pm$ 0.00 & 0.50 $\pm$ 0.00 \\
 & Multivariate cox & 0.53 $\pm$ 0.00 & 0.53 $\pm$ 0.00 & 0.53 $\pm$ 0.00 & 0.52 $\pm$ 0.00 \\
 & Multivariate tree & 0.50 $\pm$ 0.00 & 0.50 $\pm$ 0.00 & 0.50 $\pm$ 0.00 & 0.50 $\pm$ 0.00 \\
 & Univariate interaction & 0.54 $\pm$ 0.01 & 0.55 $\pm$ 0.01 & 0.55 $\pm$ 0.01 & 0.54 $\pm$ 0.01 \\
 & Univariate t-test & 0.51 $\pm$ 0.00 & 0.51 $\pm$ 0.00 & 0.50 $\pm$ 0.00 & 0.50 $\pm$ 0.00 \\
\cline{1-6}
\multirow[t]{8}{*}{0.437164} & ARDP & 0.50 $\pm$ 0.00 & 0.50 $\pm$ 0.00 & 0.50 $\pm$ 0.00 & 0.50 $\pm$ 0.00 \\
 & ITree & 0.66 $\pm$ 0.01 & 0.69 $\pm$ 0.01 & 0.67 $\pm$ 0.02 & 0.62 $\pm$ 0.02 \\
 & MOB & 0.53 $\pm$ 0.01 & 0.53 $\pm$ 0.01 & 0.52 $\pm$ 0.01 & 0.52 $\pm$ 0.01 \\
 & Multivariate cox & 0.54 $\pm$ 0.00 & 0.53 $\pm$ 0.00 & 0.53 $\pm$ 0.00 & 0.53 $\pm$ 0.00 \\
 & Multivariate tree & 0.66 $\pm$ 0.03 & 0.63 $\pm$ 0.03 & 0.57 $\pm$ 0.03 & 0.50 $\pm$ 0.01 \\
 & Univariate interaction & 0.59 $\pm$ 0.00 & 0.60 $\pm$ 0.00 & 0.59 $\pm$ 0.00 & 0.59 $\pm$ 0.01 \\
 & Univariate t-test & 0.53 $\pm$ 0.01 & 0.52 $\pm$ 0.01 & 0.52 $\pm$ 0.01 & 0.51 $\pm$ 0.01 \\
\cline{1-6}
\hline
\end{tabular}
\end{center}
\end{table}

\end{document}